\documentclass[5p]{elsarticle}

\usepackage{lineno}

\usepackage[T1]{fontenc}

\usepackage[english]{babel}
\usepackage{amssymb}
\usepackage{amsfonts}
\usepackage{amsmath}
\usepackage{txfonts}
\usepackage{graphicx}
\usepackage{ae,aecompl}
\usepackage{fancyhdr}
\usepackage{multicol}
\usepackage{layout}
\usepackage{graphicx}
\usepackage{multirow}
\usepackage{times}
\usepackage{textcomp}
\usepackage{cprotect}
\usepackage[dvipsnames]{xcolor}

\usepackage[outdir=./figures/]{epstopdf}
\usepackage[export]{adjustbox}

\usepackage{hyperref}
\hypersetup{%
    ,urlcolor=blue
    ,citecolor=blue
    ,linkcolor=blue
    }

\graphicspath{{./figures/}}

\biboptions{sort&compress}

\newif\ifAMStwofonts
\AMStwofontstrue

\def\be{\begin{equation}}
\def\ee{\end{equation}}
\def\ba{\begin{eqnarray}}
\def\ea{\end{eqnarray}}
\def\l{\left}
\def\r{\right}
\def\f{\frac}

\def\nn{\nonumber}

\begin{document}

\begin{frontmatter}

\title{Growth of non-linear structures and spherical collapse \\ in the Galileon Ghost Condensate model}

\author[1]{Noemi Frusciante}
\ead{nfrusciante@fc.ul.pt}

\author[2]{Francesco Pace}
\ead{francesco.pace@manchester.ac.uk}

\address[1]{Instituto de Astrof\'isica e Ci\^encias do Espa\c{c}o, Faculdade de Ci\^encias da Universidade de Lisboa,  
\\ Edificio C8, Campo Grande, P-1749016, Lisboa, Portugal}

\address[2]{Jodrell Bank Centre for Astrophysics, School of Natural Sciences, Department of Physics and Astronomy, \\
The University of Manchester, Manchester, M13 9PL, U.K.}

\begin{abstract}
We present a detailed study of the collapse of a spherical matter overdensity and the non-linear growth of large scale 
structures in the Galileon ghost condensate (GGC) model. This model is an extension of the cubic covariant Galileon 
(G3) which includes a field derivative of type $(\nabla_\mu\phi\nabla^\mu\phi)^2$ in the Lagrangian. We find that the 
cubic term activates the modifications in the main physical quantities whose time evolution is then strongly affected 
by the additional term. Indeed, the GGC model shows largely mitigated effects in the linearised critical density 
contrast, non-linear effective gravitational coupling and the virial overdensity with respect to G3 but still preserves 
peculiar features with respect to the standard $\Lambda$CDM cosmological model, e.g., both the linear critical density 
contrast and the virial overdensity are larger than those in $\Lambda$CDM. The results of the spherical collapse model 
are then used to predict the evolution of the halo mass function, non-linear matter and lensing power spectra. 
While at low masses the GGC model presents about 10\% fewer objects with respect to $\Lambda$CDM, at higher masses for 
$z>0$ it predicts 10\% ($z=0.5$)-20\% ($z=1$) more objects per comoving volume. Using a phenomenological approach to 
include the screening effect in the matter power spectrum, we show that the difference induced by the modifications of 
gravity are strongly dependent on the screening scale and that differences can be up to 20\% with respect to 
$\Lambda$CDM. These differences translate to the lensing power spectrum where qualitatively the largest differences 
with respect to the standard cosmological model are for $\ell<10^3$. Depending on the screening scale, they can be up 
to 25\% on larger angular scales and then decrease for growing $\ell$. These results are obtained for the best fit 
parameters from linear cosmological data for each model.
\end{abstract}


\end{frontmatter}

\section{Introduction}\label{sec:intro}
The late-time acceleration of the Universe has been confirmed by several cosmological observations 
\cite{Riess:1998cb,Perlmutter:1998np,Betoule:2014frx,Spergel:2003cb,Ade:2015xua,Aghanim:2015xee,Eisenstein:2005su,
Beutler:2011hx}. Its modelling within General Relativity (GR) is done through the cosmological constant $\Lambda$ which 
counteracts the attractive force of gravity realising the desired acceleration. The resulting model is the 
$\Lambda$-cold-dark-matter ($\Lambda$CDM) which provides an accurate picture of the Universe. However, it still 
contains a number of open theoretical issues \cite{Joyce:2014kja} which might signal the breakdown of GR. Alternative 
proposals, known as modified gravity theories (MG), suggest to modify the gravitational interaction on cosmological 
scales. The latter usually foresee the inclusion of additional degrees of freedom (dofs) 
\cite{Lue:2004rj,Copeland:2006wr,Silvestri:2009hh,Capozziello:2011et,Clifton:2011jh,Tsujikawa:2010zza,Joyce:2014kja,
Koyama:2015vza,Nojiri2017,Ferreira:2019xrr,Kobayashi2019,Ishak2019}. Among these proposals, scalar-tensor theories of 
gravity have played a prominent role as they simply add a scalar dof to the usual tensor modes of GR 
\cite{Horndeski:1974wa,Fujii:2003pa,Deffayet:2009mn,Clifton:2011jh,Tsujikawa:2010zza,Gleyzes:2014dya,Joyce:2014kja,
Koyama:2015vza,Langlois:2015cwa,Ferreira:2019xrr,Ishak2019}. For example, Horndeski theory (or Galileon theory) 
\cite{Horndeski:1974wa,Deffayet:2009mn,Kobayashi:2011nu}, is described by an action characterized by four free 
functions of the scalar field $\phi$ and its kinetic energy $X=\nabla_\mu\phi \nabla^\mu\phi$. In this theory, the 
scalar field obeys a second order Euler-Lagrange equation and a fixed form for these functions defines a model. In the 
last decade several Galileon models have been proposed 
\cite{Deffayet:2009wt,Deffayet:2010qz,DeFelice:2011bh,Gomes:2013ema,Kase:2018iwp,Albuquerque:2018ymr,Frusciante:2018aew,
Kase:2018aps} and some of them have been tested against cosmological data at linear level 
\cite{DeFelice:2011aa,Barreira:2013jma,Barreira:2014jha,Renk:2017rzu,Peirone:2017vcq,Peirone:2019aua,Giacomello:2018jfi,
Frusciante:2019puu}. The so-called  Galileon ghost condensate model (GGC) \cite{Deffayet:2010qz,Kase:2018iwp} is of 
particular interest as it is the first Galileon model to be statistically preferred by data over $\Lambda$CDM 
\cite{Peirone:2019aua}. This is due to a suppression in the low-$\ell$ tail of the Cosmic Microwave Background (CMB) 
temperature-temperature power spectrum with respect to $\Lambda$CDM and a peculiar evolution of the expansion history, 
characterised by a dark energy (DE) equation of state $w_{\rm DE}$ entering the region $(-2, -1)$ during the matter era 
without ghosts.

The GGC model possesses a screening mechanism, dubbed Vainshtein mechanism 
\cite{Vainshtein:1972sx,Nicolis:2008in,Koyama:2013paa,Kimura:2011dc,Joyce:2014kja}, which suppresses the modifications 
to gravity on Solar-System scales where GR is tested with exquisite precision~\cite{Uzan2011,Will2014}. 
The Vainshtein mechanism operates through the second derivative of the scalar field $\partial^2\phi$, dropping the 
modification to the gravity force in high-density environment. Screening mechanisms play a very important role when 
considering the formation of gravitationally bound structures: indeed, during the collapse phase the density of the 
region can be sufficiently high to significantly modify the dynamics of the scalar field. Analysis in this direction 
have been performed using the spherical collapse model for Galileon models 
\cite{Schmidt:2009yj,Kimura:2010di,Bellini:2012qn,Barreira:2013eea}. For example, in DGP braneworld gravity, the 
Vainshtein mechanism affects both force and energy conditions during collapse, in particular the conservation of the 
Newtonian total energy is violated \cite{Schmidt:2009yj} and in both DGP and cubic Galileon models an enhancement of 
structure formation is found due to a screening mechanism which is not effective until late time in the collapse 
\cite{Schmidt:2009yj,Barreira:2013eea}. 
Hence, in order to properly constrain any MG model, it is important to fully understand the impact the screening 
mechanism has on the dynamics of the scalar field at non-linear scales. This investigation is timely and relevant in 
light of new and high quality data in the non-linear regime of weak lensing and galaxy clustering from upcoming surveys, 
e.g., \textit{LSST}\footnote{\href{https://www.lsst.org/}{https://www.lsst.org/}} \citep{LSST2009,LSST2012}, 
\textit{DESI}\footnote{\href{desi.lbl.gov}{desi.lbl.gov}} \citep{DESI2016}, 
\textit{Euclid}\footnote{\href{https://www.euclid-ec.org}{https://www.euclid-ec.org}} 
\citep{Laureijs2009,Laureijs:2011gra}, 
\textit{SKA}\footnote{\href{https://www.skatelescope.org/}{https://www.skatelescope.org/}} \citep{Weltman2020}.

In this paper we aim at investigating how the effects of non-linearities in the GGC scenario change the collapse 
process of a spherical overdensity and what the role of the Vainshtein mechanism is. We will then use the results of 
the analysis of the spherical collapse model to make theoretical predictions on the abundances of halos and discuss the 
non-linear matter and lensing power spectra.

The work is organised as follows. In Section~\ref{sec:model} we introduce the GGC model and give an overview of the 
background equations. In Section~\ref{sec:linperturb} we present the linearly perturbed equations and the evolution of 
the linear matter density perturbation and its growth rate. Then, in Section~\ref{sec:nonlinperturb} we derive the 
non-linear corrections to the equations for both the scalar field and matter perturbations. The spherical collapse is 
then studied in Section~\ref{Sec:sphericalcollapse}. In Section~\ref{sec:nonlinmatterlensing} we present the 
theoretical predictions for the non-linear matter and lensing power spectra, and in Section~\ref{sec:massfunction} the 
effects of the GGC model on the abundances of halos. We finally conclude in Section \ref{sec:conclusion}.

\section{The model}\label{sec:model}

The Galileon ghost condensate (GGC) model is defined by the following action \cite{Deffayet:2010qz,Kase:2018iwp}
\be
\mathcal{S} = \int {\rm d}^4 x \sqrt{-g} \left[ \frac{M_{\rm pl}^2}{2}R + 
a_1 X+a_2 X^2+3a_3X \Box \phi \right]\,,
\label{action}
\ee
where $M_{\rm pl}$ is the Planck mass, $g$ is the determinant of the metric $g_{\mu\nu}$, $R$ is the Ricci scalar, 
$X=\nabla_\mu\phi \nabla^\mu\phi$ with $\phi$ being the scalar field and $\nabla_\mu$ the covariant derivative. 
$a_{1,2,3}$ are constants and $\Box \phi=\nabla_\mu\nabla^\mu\phi$. To the action in (\ref{action}) we add the matter 
action ${\cal S}_{\rm M}$, for which we consider perfect fluids minimally coupled to gravity.

Varying the total action with respect to the metric and $\phi$, we obtain the corresponding field equations. We 
consider for the background the flat Friedmann-Lema{\^i}tre-Robertson-Walker (FLRW) line element, given by
\be
 {\rm d}s^2=-{\rm d}t^2 + a^2(t)\gamma_{ij}{\rm d}x^i {\rm d}x^j\,,
\ee 
where $a(t)$ is the scale factor and $\gamma_{ij}$ is the spatial metric. Following Ref.~\cite{Kase:2018iwp}, we 
introduce the dimensionless variables
\be\label{dimensionless_functions}
x_1 = -\f{a_1\dot{\phi}^2}{3M_{\rm pl}^2 H^2}\,,\quad 
x_2 = \f{a_2\dot{\phi}^4}{M_{\rm pl}^2 H^2}\,, \quad 
x_3 = \f{6a_3\dot{\phi}^3}{M_{\rm pl}^2H}\,,
\ee
where $H=\dot{a}/a$, and a dot represents the derivative with respect to the cosmic time $t$. Using these definitions 
we can write the field equations in the background as a dynamical system:
\ba\label{system}
 && x_1^{\prime} = 2x_1(\epsilon_\phi-h)\,,\qquad 
    x_2^{\prime} = 2x_2(2\epsilon_\phi-h)\,, \\
 && x_3^{\prime} = x_3(3\epsilon_\phi-h)\,,\qquad
    \Omega_{\rm r}^{\prime} = -2\Omega_{\rm r}(2+h)\,,
\ea
where $\Omega_{\rm r}=\rho_{\rm r}/(3M_{\rm pl}^2 H^2)$ is the dimensionless density parameter for radiation, 
$\epsilon_{\phi}=\ddot{\phi}/(H \dot{\phi})$, $h=\dot{H}/H^2$, and a prime is defined as the derivative with respect to 
${\cal N}=\ln{a}$. Given the length of the expressions for $\epsilon_{\phi}$ and $h$ we refer the reader to Eqs.~(4.16) 
and (4.17) in Ref.~\cite{Kase:2018iwp} (with $x_4=0$). From the Friedmann equation we also have 
\be
\Omega_{\rm c} + \Omega_{\rm b} + \Omega_{\rm r} + \Omega_{\rm DE} = 1\,,
\ee
where $\Omega_{\rm b,c} = \rho_{\rm b,c}/(3M_{\rm pl}^2 H^2)$ are the density parameter for the baryons (b) and cold 
dark matter (c), respectively, and 
\be\label{consteq}
 \Omega_{\rm DE} = x_1 + x_2 + x_3\,,
\ee
is the DE density parameter. 
Eq.~(\ref{consteq}), evaluated today, can be used to reduce the number of free parameters of the model, leaving the 
model with two additional parameters out of three compared to $\Lambda$CDM, i.e.,
\be
 x_2^{(0)}=\Omega_{\rm DE}^{(0)}- x_1^{(0)}- x_3^{(0)}\,.
\ee

The GGC model allows for a de Sitter fixed point free from ghost instability. The presence of $x_2 \ne 0$ prevents the 
model from reaching a tracker solution. The latter would be characterised by $w_{\rm DE}=-2$ during the matter era, 
while the $X^2$ term allows to temporally enter the region $-2< w_{\rm DE} < -1$~\cite{Kase:2018iwp}. This property 
allows the model to be observationally favoured over $\Lambda$CDM~\cite{Peirone:2019aua}.

\section{Linear density perturbations}\label{sec:linperturb}

Let us consider the linear perturbed line element on the flat FLRW background:
\be
{\rm d}s^2 = -\left(1+2\Psi \right) {\rm d}t^2 + 
a^2(t) \left(1-2\Phi \right) \gamma_{ij} {\rm d}x^i {\rm d}x^j\,,
\ee
where $\Psi(t,x_i)$ and $\Phi(t,x_i)$ are the gravitational potentials. In Fourier space, for MG models with one extra 
scalar dof we can write the following equations which generalise the standard general relativistic Poisson and lensing 
equations \cite{Amendola:2007rr,Bertschinger:2008zb,Pogosian:2010tj}:
\ba
\label{mudef}
&& -k^2\Psi = 4\pi G_{\rm N} a^2\mu^{\rm L}(a,k)\rho_{\rm m}\delta_{\rm m}\,, \\
&& -k^2(\Psi+\Phi) = 8\pi G_{\rm N} a^2\Sigma^{\rm L}(a,k)\rho_{\rm m} \delta_{\rm m}\,,
\ea
where $G_{\rm N}^{-1}=8\pi M_{\rm pl}^2$ is the Newtonian gravitational constant, $k$ is  the comoving wavenumber, 
$\rho_{\rm m} \delta_{\rm m} = \sum_{i} \rho_i \delta_i$ is the total matter density perturbation (where 
$i\in [{\rm r}, {\rm b}, {\rm c}]$). 
The dimensionless quantities $\mu^{\rm L}$ and $\Sigma^{\rm L}$ characterise the effective gravitational couplings at 
linear order felt by matter and light, respectively. The GR limit is recovered when both 
$\mu^{\rm L}=\Sigma^{\rm L}=1$. Applying the quasi-static approximation (QSA) \footnote{In the QSA, time derivatives of 
the perturbed quantities can be neglected compared with their spatial derivatives. We note that the validity of the QSA 
for the Horndeski class of models has been proved to be a valid assumption within the scalar field's sound horizon for 
$k>0.001~h/{\rm Mpc}$ \cite{Peirone:2017ywi,Frusciante:2018jzw}. We have verified that indeed this is the case for the 
GGC model.} \cite{Boisseau:2000pr,DeFelice:2011hq} for perturbations inside the scalar field's sound horizon 
\cite{Sawicki:2015zya} to the model in action~(\ref{action}), it follows that \cite{Kase:2018iwp}
\be\label{eq:mulinear}
\mu^{\rm L}(a)=\Sigma^{\rm L}(a) = 1+\frac{x_3^2}{Q_{\rm s} c_{\rm s}^2 (2-x_3)^2}\,,
\ee
where
\begin{align}
Q_{\rm s} = & \frac{3(4x_1+8x_2+4x_3+x_3^2)}{(2-x_3)^2}\,,\\
c_{\rm s}^2 = & \frac{2(1+3\epsilon_{\phi})x_3 - x_3^2 - 4h - 
6(\Omega_{\rm c} + \Omega_{\rm b})-8\Omega_{\rm r}}
{3(4x_1+8x_2+4x_3+x_3^2)}\,.
\end{align}
To avoid ghosts and Laplacian instabilities, we require that both $Q_{\rm s}$ and the speed of propagation of the 
scalar modes $c_{\rm s}^2$ are positive. Then, for $x_3 \neq 0$, $\mu^{\rm L}$ and $\Sigma^{\rm L}$ are larger than 1. 
Since $\mu^{\rm L}=\Sigma^{\rm L}$, there is no gravitational slip ($\Psi=\Phi$).

For sub-horizon perturbations, the matter density $\delta_{\rm m}$ approximately obeys the linear equation
\be\label{eqn:linpert}
\delta_{\rm m}^{\prime\prime} + \l(2+\f{H^\prime}{H}\r) \delta_{\rm m}^\prime
- \f{3}{2}\Omega_{\rm m} \mu^{\rm L}(a) \delta_{\rm m}=0\,,
\ee
where we have used Eq.~(\ref{mudef}) to replace $\Psi$ in favour of $\delta_{\rm m}$.

We solve the equation above by setting the initial conditions (ICs) as follows: $a_i=0.01$, $\delta_{{\rm m},i} = a_i$ 
and $\delta^\prime_{\rm m,i} =a_i$, which correspond to the matter dominated era solution. The model and cosmological 
parameters of GGC are listed in Tab.~\ref{tab:bestvalues} and they correspond to the cosmological constraints obtained 
with Planck data \cite{Peirone:2019aua}. For reference we also include the parameters for other two models: the 
$\Lambda$CDM model (for the constraints we refer to \cite{Peirone:2019aua}) and the Cubic Galileon model (G3) 
\cite{Deffayet:2009wt}. We decided to use this model for comparison because it can be obtained from GGC by setting 
$x_2=0$~\footnote{Let us note that the analysis for G3 we show in this work has been the subject of several papers in 
the past~\cite{Kimura:2010di,Bellini:2012qn,Barreira:2013eea}. That is why we do not explicitly rewrite the 
corresponding equations but we prefer to refer the reader to these papers for a detailed discussion.}. Given this 
property, the G3 model shows a tracker solution $H^2\phi^\prime=$const \cite{DeFelice:2010pv}. The values of the 
cosmological parameters we use for G3 are from the constraints in Ref. \cite{Peirone:2017vcq}. In this work we decided 
to use the constrain values of the cosmological/model parameters for each model because we want to make theoretical 
predictions that are as close as possible to what we can actually expect from observations.

\begin{figure}[t!]
 \includegraphics[width=.46\textwidth]{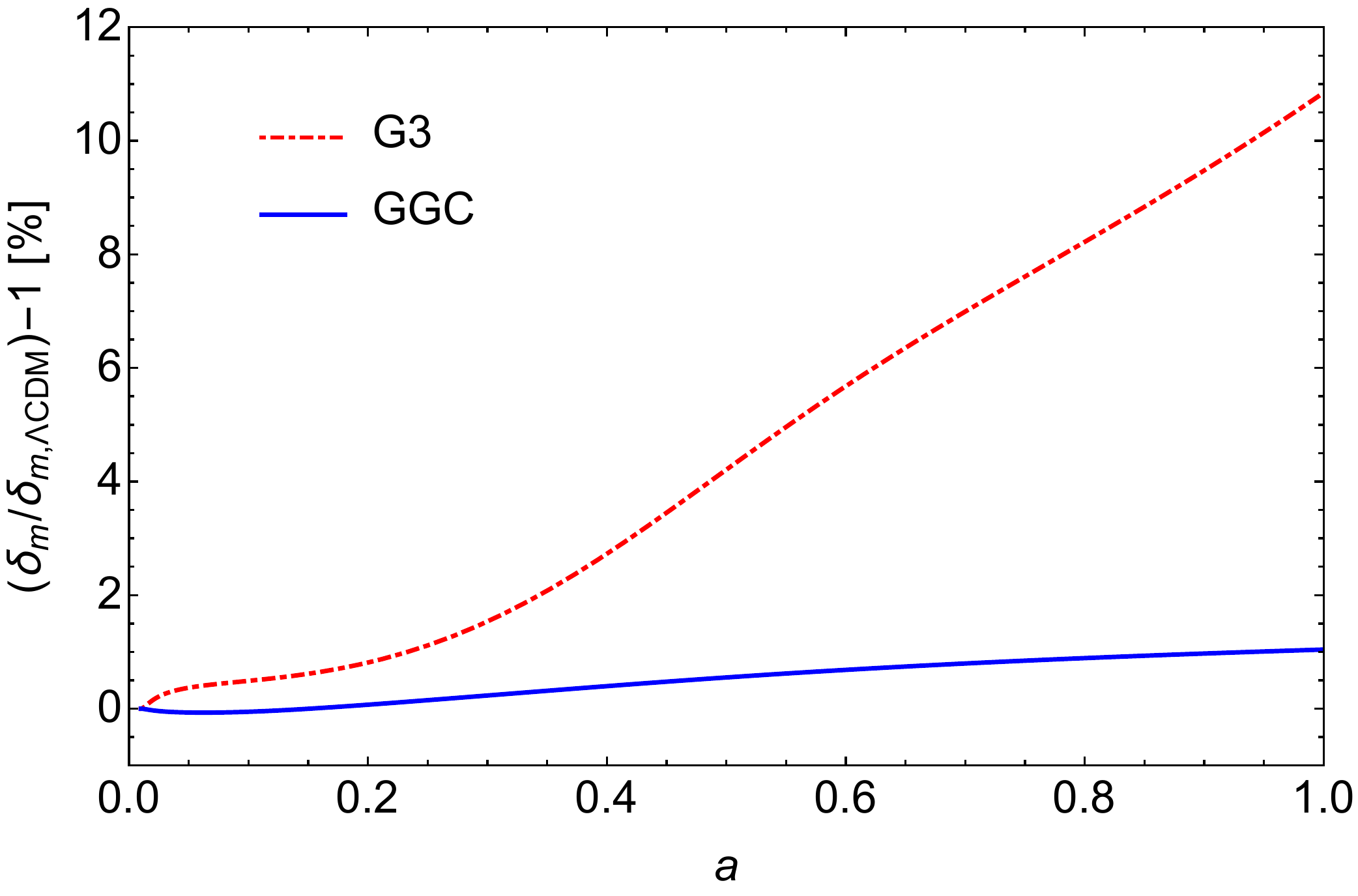}
 \includegraphics[width=.46\textwidth]{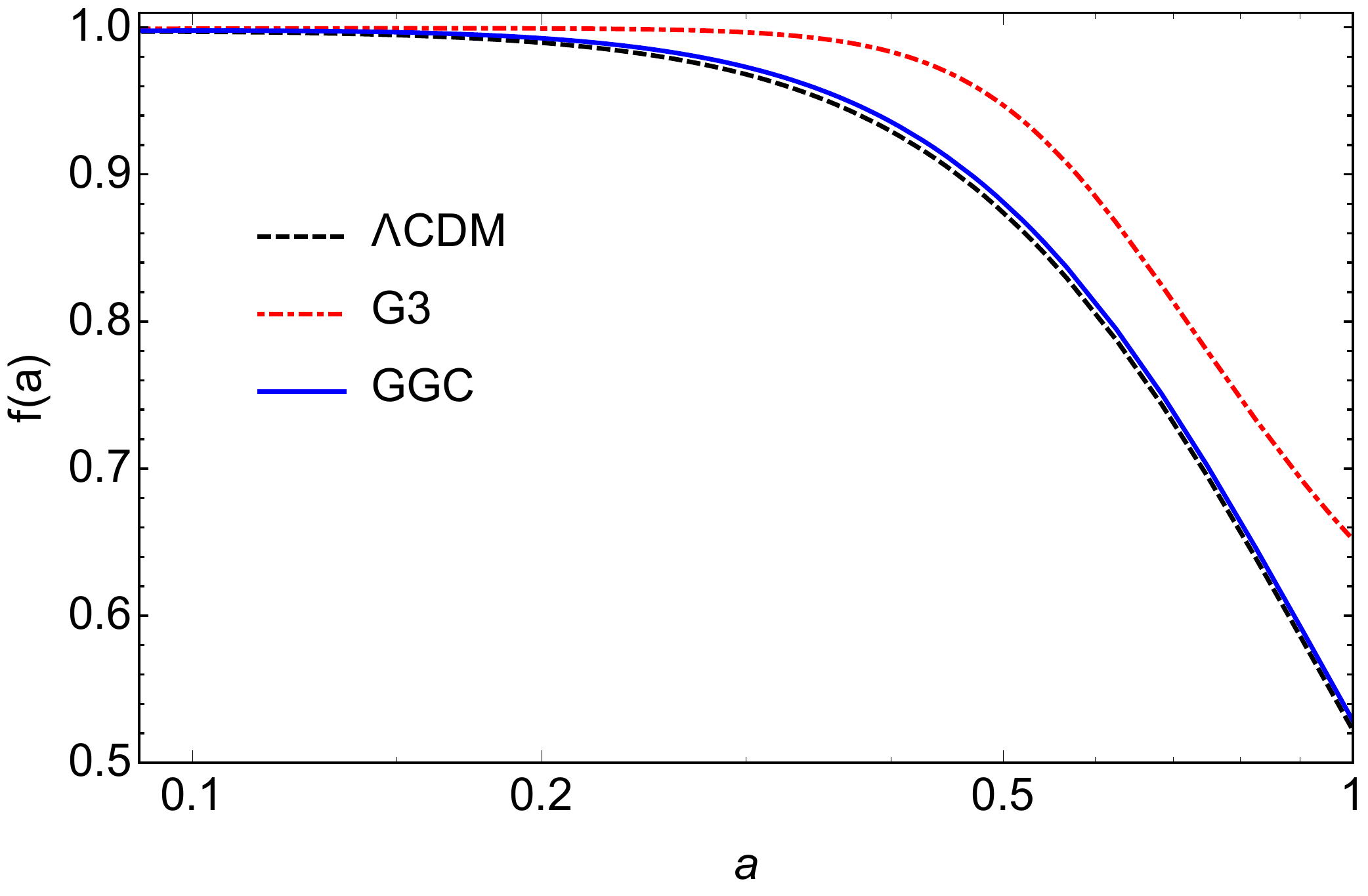}
 \caption{Top panel: Time evolution of the relative difference in percentage of the linear matter density perturbation 
 in the GGC (blue solid line), G3 (red dot-dashed line) with respect to the $\Lambda$CDM. Bottom panel: Evolution of 
 the linear growth rate for GGC, G3 and $\Lambda$CDM (black dashed line). The model and cosmological parameters are 
 shown in Tab.~\ref{tab:bestvalues}.}
 \label{fig:linear}
\end{figure}

In the top panel of Fig.~\ref{fig:linear}, we show the relative difference in the evolution of the linear matter 
density perturbation $\delta_{\rm m}$ with respect to $\Lambda$CDM for both the GGC and the G3 models. The relative 
difference is very small at early times for both models. In the case of GGC, it remains smaller than $1\%$ throughout 
its growth history, while for G3 it reaches $11\%$ at present. Such modifications with respect to the $\Lambda$CDM 
model are due in both cases to a modified expansion history and to $\mu^{\rm L} \neq 1$ at later times, while the large 
difference between the GGC and the G3 is only due to the presence of $x_2 \neq 0$ in GGC.

Modifications with respect to $\Lambda$CDM can be also spotted in the linear growth rate $f(a)$, which is a derived 
quantity defined as
\be
f(a)=\f{\mathrm{d}\ln{\delta_{\rm m}}}{\mathrm{d}\ln{a}}\,.
\ee 
We show its evolution in the bottom panel of Fig.~\ref{fig:linear} for the two Galileon models and we compare them to 
$\Lambda$CDM. The growth rates in both Galileon models become larger than $\Lambda$CDM as soon as the Universe exits 
the matter dominated era. Appreciable differences in the case of GGC are around $a=0.2$, being the time at which 
$\delta_{\rm m}({\rm GGC})$ starts to be larger than that of $\Lambda$CDM. For GGC, the linear growth rate $f$ is 
enhanced with respect to the standard model until $a \gtrsim 0.6$, while at earlier times the difference is negligible. 
In G3 differences arise a bit earlier because a 0.5\% difference in $\delta_{\rm m}$ is already present (see upper 
panel). A large enhanced modification is then present up to the present time.

\begin{table}[t!]
\centering
\begin{tabular}{|c|c|c|c|c|c|c|}
\hline
Model & $\sigma_8^{(0)}$ &$H_0$ & $\Omega_{\rm m}^{(0)}$ & $x_1^{(0)} $ & $x_2^{(0)} $& $x_3^{(0)} $ \\
\hline
$\Lambda$CDM  & 0.83 &70 & 0.31 & -- &--&--\\
\hline
G3 & 0.93 & 73.9  &0.27& --&--&-- \\
\hline
GGC & 0.87 & 70& 0.28 &-1.26&1.64&0.34 \\
\hline
\end{tabular}
\caption{Present day values for the amplitude of the linear matter power spectrum at 8 $h^{-1}$Mpc, $\sigma_8^{(0)}$, 
the Hubble parameter $H_0$ in units of  km\,s$^{-1}$Mpc$^{-1}$, the matter density $\Omega_m^{(0)}$ and the $x_i^{(0)}$ 
parameters. They correspond to the maximum likelihood values obtained with Planck data for $\Lambda$CDM and GGC in 
Ref.~\cite{Peirone:2019aua}, and the mean values of G3 obtained with Planck data in Ref.~\cite{Peirone:2017vcq}.}
\label{tab:bestvalues}
\end{table}

\section{Non-linear  perturbations}\label{sec:nonlinperturb}
We will now investigate the evolution of the metric and the scalar field perturbations on small scales, where second 
order, non-linear perturbations are no longer negligible. Let us consider the perturbation of the scalar field: 
$\phi(t,x_i)=\phi(t)+\delta\phi(t,x_i)$ and along with the QSA we will also neglect terms that are suppressed by the 
Newtonian potentials and their first spatial derivatives. 

Then, the time-time component of the GGC equation gives
\be \label{tteq}
\f{\partial^2\Phi}{a^2}=4\pi G_{\rm N}\rho_{\rm m}\delta_{\rm m} + 24\pi G_{\rm N}a_3X\f{\partial^2\delta\phi}{a^2}\,,
\ee
where the derivatives are with respect to spatial components, and the equation for the scalar field reads
\ba
-3a_3\dot{\phi}^2\f{\partial^2\Psi}{a^2} & = & \l[-a_1-2a_2X+6a_3(\ddot{\phi}+2H\dot{\phi})\r]
\f{\partial^2\delta\phi}{a^2}\nn\\
& + & 3a_3\l[\l(\f{\partial^2\delta\phi}{a^2}\r)^2-\l(\f{\partial_i\partial_j\delta\phi}{a^2}\r)^2\r]
\,,
\ea
where $\l(\partial_i\partial_j\delta\phi\r)^2=(\partial_i\partial_j\delta\phi)(\partial^i\partial^j\delta\phi)$ and 
indexes are raised with the metric $\gamma_{ij}$: $\partial^i=\gamma^{ij}\partial_j$.

At the non-linear level, the relation $\Phi=\Psi$ is still valid, so we can combine the above equations to get
\be\label{fieldeqnonlinear}
\f{\partial^2\delta\phi}{a^2} + \lambda^2\l[\l(\f{\partial^2\delta\phi}{a^2}\r)^2 - \l(\f{\partial_i\partial_j\delta\phi}{a^2}\r)^2\r] = 
-4\pi G_{\rm N}\zeta\rho_{\rm m}\delta_{\rm m}\,,
\ee
where
\ba
\lambda^2(a) = \f{12a_3\dot{\phi}^2}{M_{\rm pl}^2H^2c_{\rm s}^2Q_{\rm s}(2-x_3)^2}\,, \qquad 
\zeta(a)= \lambda^2\dot{\phi}^2\,.
\ea

Let us consider a spherically symmetric density perturbation. Then, Eq.~(\ref{fieldeqnonlinear}) becomes
\be
\f{1}{r^2} \f{\mathrm{d}}{\mathrm{d}r} \l(r^2\f{\mathrm{d}\delta\phi}{\mathrm{d}r}\r) - 
\f{2\lambda^2}{r^2}\f{\mathrm{d}}{\mathrm{d}r}\l[r\l(\f{\mathrm{d}\delta\phi}{\mathrm{d}r}\r)^2\r] = 
-4\pi G_{\rm N}\zeta\rho_{\rm m}\delta_{\rm m}\,.
\ee
Defining the mass enclosed in a sphere of radius $r$ as  
\be
m(r)=4\pi \int_0^r {r^{\prime}}^2 \rho_{\rm m}(r^\prime) \delta_{\rm m}(r^{\prime})\mathrm{d}r^{\prime}\,,
\ee
we can integrate Eq.~(\ref{fieldeqnonlinear}) and obtain
\be
r^2\f{\mathrm{d}\delta\phi}{\mathrm{d}r} - 2\lambda^2r\l(\f{\mathrm{d}\delta\phi}{\mathrm{d}r}\r)^2 = 
-G_{\rm N}\zeta m(r)\,.
\ee
We can now evaluate its solution, which reads
\be\label{fieldsol}
\f{\mathrm{d}\delta\phi}{\mathrm{d}r} = \f{r_{\rm V}}{4\lambda^2} 
\l[\f{r}{r_{\rm V}} \l(1-\sqrt{1+\f{r_{\rm V}^3}{r^3}}\r)\r]\,,
\ee
where $r_{\rm V}$ is the Vainshtein radius of the enclosed mass perturbation and it is defined as
\be
 r_{\rm V}^3 = 8G_{\rm N} m(r)\lambda^2\zeta = 
 \f{32G_{\rm N} m(r)x_3^2}{[Hc_{\rm s}^2Q_{\rm s}(2-x_3)^2]^2}\,.
\ee

\begin{figure}[t!]
\includegraphics[width=.46\textwidth]{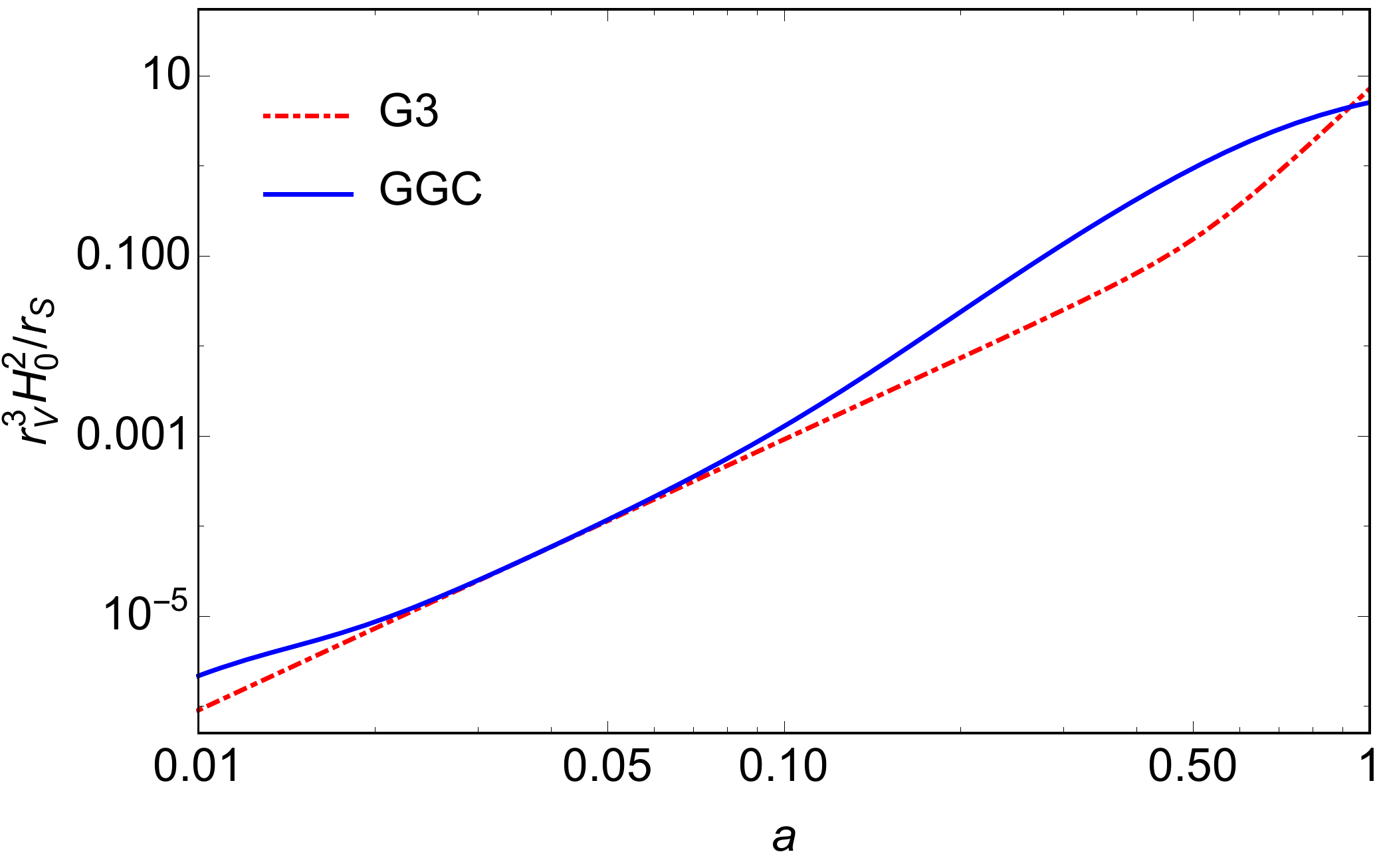}
\caption{Time evolution of the Vainshtein radius for the GGC (blue solid line) and G3 (red dot-dashed line) models. We 
have defined $r_{\rm s} = 2G_{\rm N}M$ as the Schwarzschild radius. The parameters for the GGC and the G3 models have 
been chosen according to the cosmological constraints in \cite{Peirone:2019aua} and \cite{Peirone:2017vcq}, 
respectively.}
\label{fig:Vradius}
\end{figure}

It then depends on the mass distribution in the sphere and on the parameters of the model. In particular, it is 
non-vanishing as long as $x_3 \neq 0$. For a point source, $r_{\rm V} = 2.07\times10^2(M/M_\odot)^{1/3}$ pc, where we 
have used the maximum likelihood values for the present day parameters $H_0, x_3, x_2, x_1$ obtained in 
\cite{Peirone:2019aua} with Planck data. The corresponding Vainshtein radius for G3 is $r_{\rm V} = 
2.24\times10^2(M/M_\odot)^{1/3}$ pc where we have used the constraints obtained in \cite{Peirone:2017vcq} (see 
Tab.~\ref{tab:bestvalues}). We note that the Vainshtein radius at the present time for the GGC is smaller than the one 
for the G3. In Fig.~\ref{fig:Vradius} we compare the time evolution of the Vainshtein radius for both the GGC and the 
G3 models. For both models, at early times, its value is very small which means that the screening mechanism works only 
on very small scales. At this time the Vainshtein radius for the GGC is slightly larger than the G3 one and afterwards 
they become equal. As the Universe expands, the GGC radius increases and its value is higher if compared to G3 in the 
time range $0.07<a<0.9$. Only at present time we notice a change of trend, the Vainshtein radius of the GGC decreases 
with respect to the G3 according to the estimated values we presented before.

According to Eq.~(\ref{fieldsol}), well outside the Vainshtein radius, the derivative of the scalar field perturbation 
is proportional to the Newtonian potential and it corresponds to the linear solution.

If we consider a top-hat profile for the density field, we get that 
$\mathrm{d}\delta\phi/\mathrm{d}r \propto r$ for $r<R$ where $R$ is the radius of the sphere of mass $m(R)=M$. Then, 
Eq.~(\ref{fieldeqnonlinear}) reduces to
\be
\f{\partial^2\delta\phi}{a^2} - \f{2\lambda^2}{3}\l(\f{\partial^2\delta\phi}{a^2}\r)^2 = 
-4\pi G_{\rm N}\zeta \rho_{\rm m}\delta_{\rm m}\,.
\ee
At $r=R$ the equation above can be solved for $\partial^2\delta\phi/a^2$ and we find
\be
\f{\partial^2\delta\phi}{a^2} = 8\pi G_{\rm N}\rho_{\rm m}\zeta
\l(\f{R}{R_{\rm V}}\r)^3\l[1-\sqrt{1+\f{R_{\rm V}^3}{R^3}}\r]\delta_{\rm m}\,,
\ee
where $R_{\rm V}^3=8G_{\rm N}\lambda^2\zeta \delta M$\, and $\delta M$ is the total mass of the density perturbation 
$\rho_{\rm m}\delta_{\rm m}$. Now we can compute a modified Poisson equation from Eq.~(\ref{tteq}) which includes 
non-linear corrections and it reads
\be\label{nonlinearpoisson}
\f{\partial^2 \Psi}{a^2} = 4\pi G_{\rm N} \mu^{\rm NL}(a,R)\,\rho_{\rm m}\delta_{\rm m}\,,
\ee
where the non-linear effective gravitational coupling is
\ba\label{muNL}
\mu^{\rm NL}(a,R) = 
1 + 2\l(\mu^{\rm L}-1\r)\l(\f{R}{R_{\rm V}}\r)^3
\l(\sqrt{1+\f{R_{\rm V}^3}{R^3}}-1\r)\,.\,
\ea
In the limit $R\rightarrow 0$ the above expression reduces to unity approaching GR, while for $R\gg R_{\rm V}$ we 
recover the linear result $\mu^{\rm NL}\rightarrow \mu^{\rm L}$, showing how the Vainshtein screening mechanism works.

In the next Section we will show the evolution of $\mu^{\rm NL}$ for a collapsing sphere and we will compare it with 
the linear effective gravitational coupling. Finally, because in the non-linear regime $\Phi=\Psi$ is still valid, we 
can deduce $\Sigma^{\rm NL}=\mu^{\rm NL}$. This information will be used when computing the non-linear lensing power 
spectrum.

\section{Spherical collapse model}\label{Sec:sphericalcollapse}

The spherical collapse process is the simplest model for the formation of non-linear gravitationally bound structures. 
It is characterised by the turnaround phase, during which the amplitude of the spherical perturbation in the expansion 
phase reaches a sufficient large value such that the gravitational force prevents the sphere from an infinite 
expansion. It is then followed by the proper collapse phase, i.e., when the sphere reaches its maximum radius at the 
turnaround, $R_{\rm ta}$, the overdensity starts to collapse. While the mathematical model implies the collapsing 
sphere to reduce to a point, in reality this does not happen, as virialization takes place 
\citep{Lahav:1991wc,Pace2019b} and the system satisfies the Virial theorem. In the following we assume that during the 
evolution the matter distribution remains with a top-hat profile.
 
The non-linear evolution equation for the matter overdensity is \citep{Pace2010,Pace2017a}
\be\label{nonlinearoverdensityeq}
\ddot{\delta}_{\rm m} + 2H\dot{\delta}_{\rm m} - \f{4}{3}\f{\dot{\delta}_{\rm m}^2}{1+\delta_{\rm m}} = 
\l(1+\delta_{\rm m}\r)\f{\partial^2\Psi}{a^2}\,.
\ee
We can use Eq.~(\ref{nonlinearpoisson}) to eliminate the metric potential. Thus it is clear that the evolution of 
$\delta_{\rm m}$ is modified with respect to GR by $\mu^{\rm NL}$.

Assuming that the total mass inside $R$ is conserved during the collapse, we have
\be
M = \f{4\pi}{3}R^3\rho_{\rm m}(1+\delta_{\rm m})=\text{const} \,,
\ee
from which we can derive the equation of the evolution of the radius after differentiating it with respect to time and 
using Eq.~(\ref{nonlinearoverdensityeq}). Then, we have 
\be\label{collapseR}
 \f{\ddot{R}}{R} = H^2 + \dot{H} - 
 \f{4\pi G_{\rm N}}{3}\mu^{\rm NL}\,\rho_{\rm m}\delta_{\rm m}\,,
\ee
which is composed by a background term ($H^2+\dot{H}$) and a gravitational one 
($\propto \mu^{\rm NL}\rho_{\rm m}\delta_{\rm m}$). We can numerically solve the above equation as follows. 
As standard procedure, we introduce the variable
\be\label{defy}
 y=\f{R}{R_i} - \f{a}{a_{i}}\,,
\ee
where $R_{i}$ is the initial radius of the perturbation and $a_{i}$ is the initial scale factor. Thus 
Eq.~(\ref{collapseR}) reads
\be\label{yeq}
y^{\prime\prime} = -\f{H^\prime}{H}y^\prime + \l(1+\f{H^\prime}{H}\r)y - 
\f{\Omega_{\rm m}}{2}\mu^{\rm NL}\delta_{\rm m}\l(y-\f{a}{a_{i}}\r)\,.
\ee
In order to specify the evolution of $\mu^{\rm NL}(a,R)$, we also use
\ba\label{RRV3}
\l(\f{R}{R_{\rm V}}\r)^3 & = & \f{1}{4\Omega_{\rm m}H^2\lambda^2\zeta }\f{1}{\delta_{\rm m}} = 
\f{x_3^2}{16\Omega_{\rm m} (\mu^{\rm L}-1)^2 } \f{1}{\delta_{\rm m}}\,,
\ea
which can be easily computed from the mass conservation and the definition of Vainshtein radius. Eq.~(\ref{RRV3}) thus 
shows the relation between the Vainshtein radius and the collapsing overdensity, which holds as long as $x_3\neq 0$.

In order to solve Eq.~(\ref{yeq}) numerically, we consider the initial conditions such that the collapse time is 
$a_{\rm collapse}=1$. It follows\footnote{From Eq.~(\ref{defy}) at initial time one gets $y_i=0$ and 
$y^\prime_i=-\delta_{{\rm m},i}^\prime/(3(1+\delta_{{\rm m},i}))$. Assuming that the density perturbation grows 
linearly during matter dominated era, we can use $\delta_{\rm m}\propto a$ and 
$\delta_{\rm m}^\prime \propto \delta_{\rm m}$, thus one gets $y^\prime_i = -\delta_{{\rm m},i}/3$ (see 
Ref. \cite{Bellini:2012qn}).}: $a_i=6.66 \times 10^{-6}$, $y_i = 0$ and $y^\prime_i = -\delta_{{\rm m},i}/3$, where 
$\delta_{{\rm m},i}$ is the initial density obtained from linear theory in the matter dominated era assuming the 
collapse ($R=0$) at $a_{\rm collapse}=1$. Finally, we write the overdensity as
\be
 \delta_{\rm m} = (1+\delta_{{\rm m},i})\l(1+\f{a_{i}}{a}y\r)^{-3}-1\,,
\ee
which follows from matter conservation.

\begin{figure}[ht!]
\includegraphics[width=.46\textwidth]{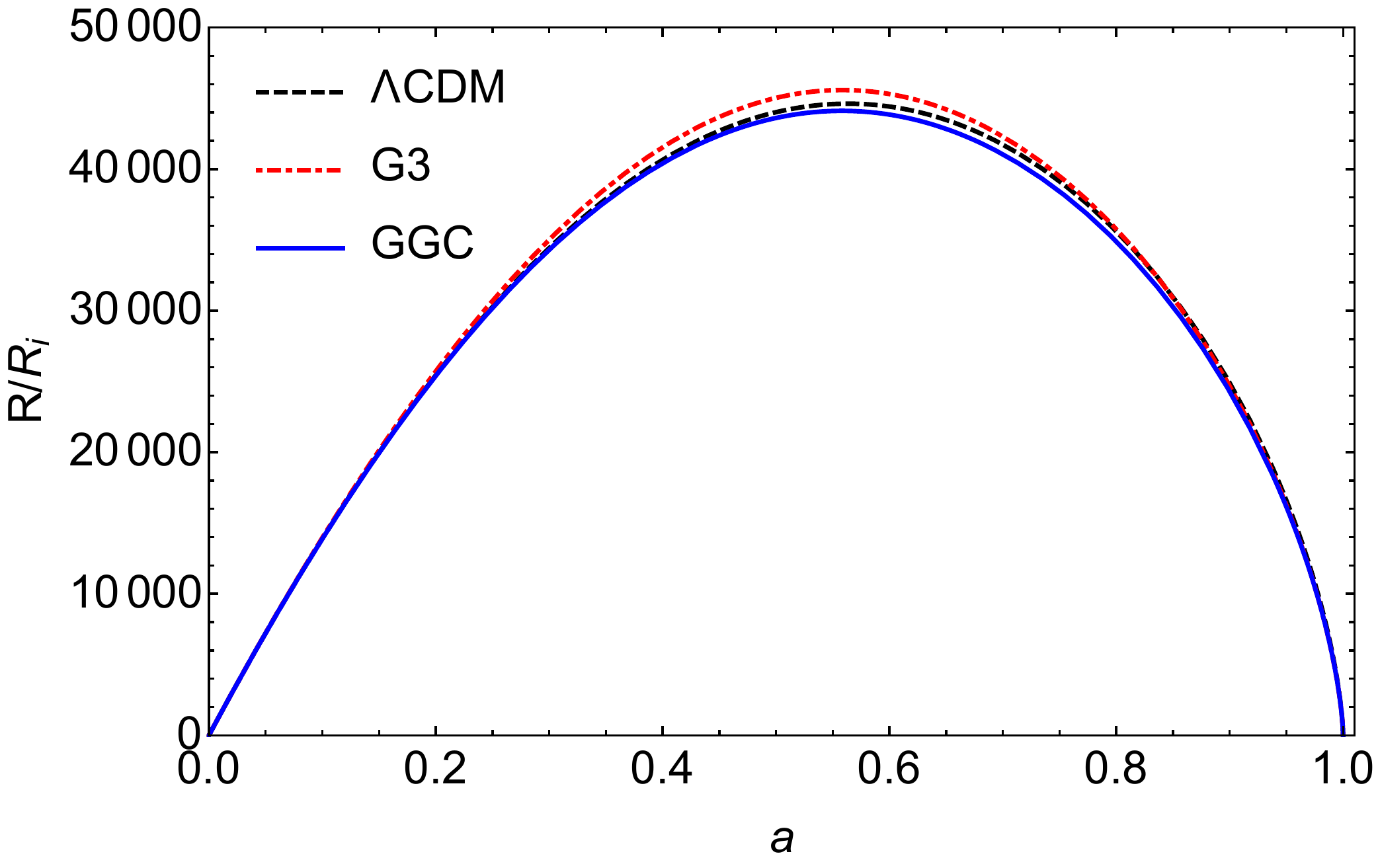}
\caption{Time evolution of $R/R_{i}$ for the GGC (blue solid line) and G3 (red dot-dashed line) and $\Lambda$CDM (black 
dashed line) models. The initial overdensity $\delta_{{\rm m},i}$ for the models are: 
$\delta_{{\rm m},i}(\Lambda {\rm CDM})=12.3\times 10^{-5}$, $\delta_{{\rm m},i}({\rm GGC})=13.2 \times 10^{-5}$ and 
$\delta_{{\rm m},i}(\rm G3)=7.3 \times 10^{-5}$. The model and cosmological parameters are shown in 
Tab.~\ref{tab:bestvalues}.}
\label{fig:RRi} 
\end{figure}

\begin{table}[t!]
 \centering
 \begin{tabular}{|c|c|c|c|}
  \hline
  & $\Lambda$CDM & G3 & GGC \\
  \hline
  $\delta_{\rm c}$ & 1.675 & 1.738 &1.708 \\
  \hline
  $\Delta_{\rm vir}$ & 333.1 & 317.6 & 342.8 \\
  \hline
  $a_{\rm ta}$ & 0.563 & 0.558 & 0.558 \\
  \hline
  $a_{\rm vir}$ & 0.922 & 0.916 & 0.919 \\
  \hline
  $R_{\rm ta}/R_i$ & 44621.5 & 45583.9 & 44114.4 \\
  \hline
  $R_{\rm vir}/R_i$ & 21639.3 & 22285.0 & 21434.5 \\
  \hline
 \end{tabular}
 \caption{Physical quantities characterising the spherical collapse at the present time for the GGC model in comparison 
 with $\Lambda$CDM and G3. The parameters of the models used to obtain these results are in Tab.~\ref{tab:bestvalues}.}
 \label{tab:quantities}
\end{table}

In Fig.~\ref{fig:RRi} we show the solution of Eq.~(\ref{collapseR}) for the three models when the collapse time is set 
at the present time. We note that modifications with respect to the $\Lambda$CDM model are present during the collapse 
phase for both Galileon models. However, the modification introduced by $x_2$ makes the dynamics of the collapse for 
the GGC quite different from that of the G3. The $\Lambda$CDM model indeed is in between the two Galileon models. This 
is understood by noticing that there is the following hierarchy for the initial overdensities: 
$\delta_{{\rm m},i}({\rm G3})<\delta_{{\rm m},i}(\Lambda{\rm CDM})<\delta_{{\rm m},i}({\rm GGC})$. 
This translates to an opposite hierarchy for the radii, as a larger overdensity implies an earlier collapse and 
therefore a smaller radius. The GGC model has a turn-around radius smaller than both the $\Lambda$CDM and the G3 as 
shown in Tab.~\ref{tab:quantities}. The latter, instead, shows the larger one. The turn-around phase takes place at the 
same time for both the Galileon cosmologies, $a_{\rm ta} \approx 0.558$ while in $\Lambda$CDM it is slightly delayed, 
$a_{\rm ta}\approx 0.567$.

\begin{figure}[t!]
\includegraphics[width=.46\textwidth]{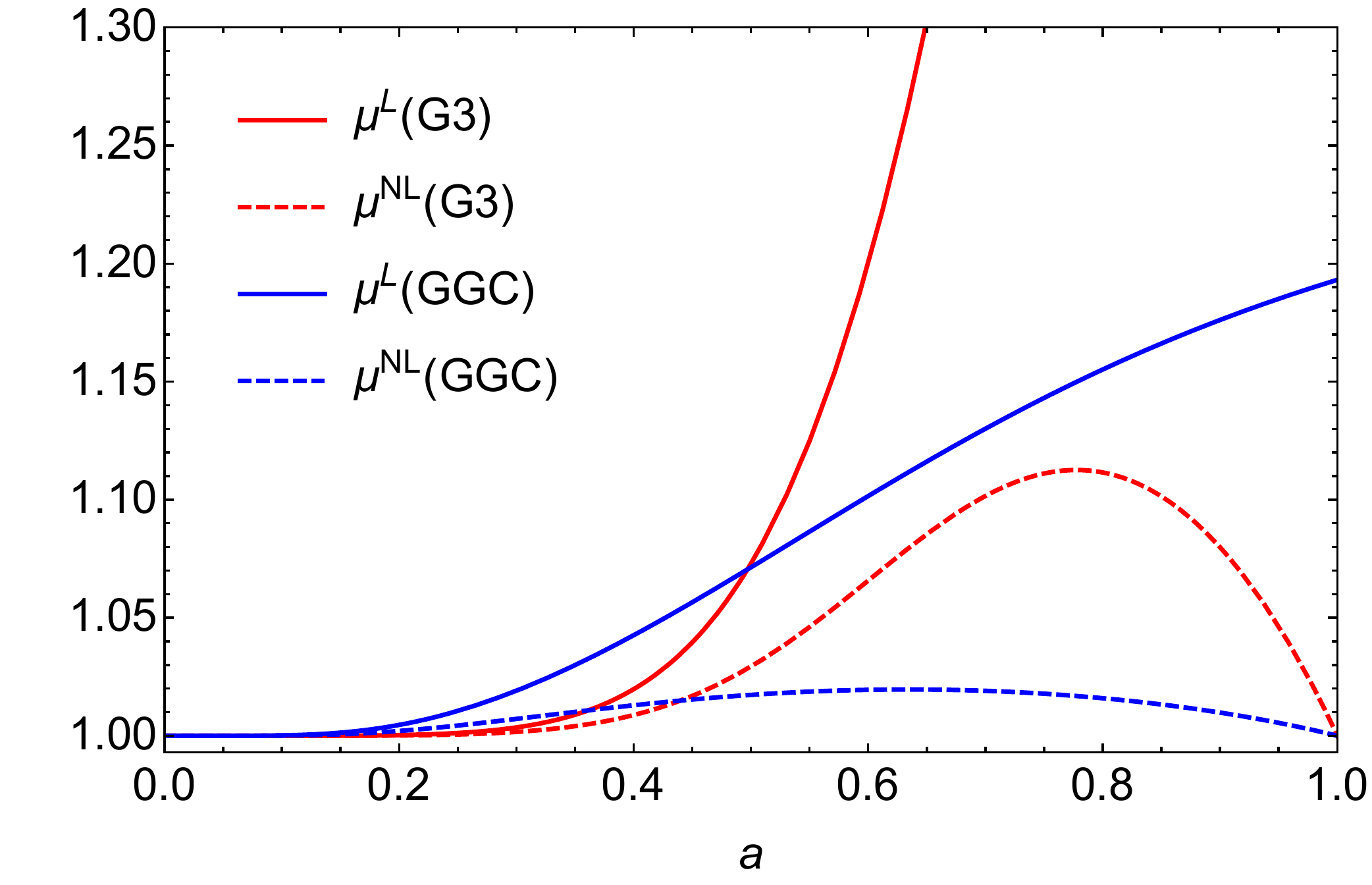}
\caption{Time evolution of $\mu^{\rm L}$ (solid lines) and $\mu^{\rm NL}$ (dashed lines) for the GGC (blue) and the G3 
(red) models. The models' parameters for the GGC and the G3 have been chosen according to the cosmological constraints 
in Refs.~\cite{Peirone:2019aua} and~\cite{Peirone:2017vcq}, respectively.}
\label{fig:muLvsmuNL} 
\end{figure}

\begin{figure}[ht!]
\includegraphics[width=.46\textwidth]{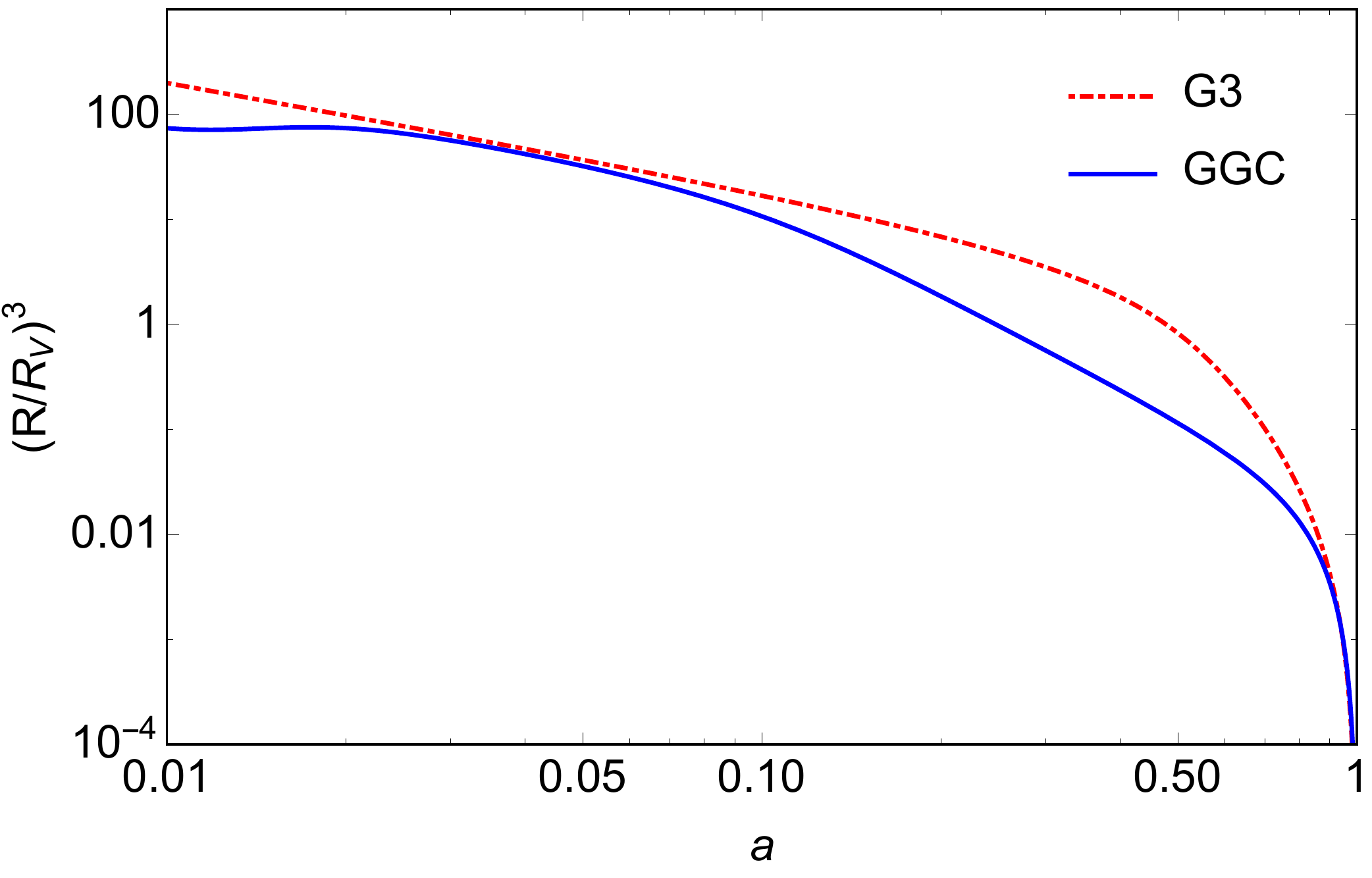}
\caption{Time evolution of $R/R_{\rm V}$ for the GGC (blue solid line) and the G3 (red dot-dashed line) models for a 
matter overdensity collapsing at the present time. The models' parameters for the GGC and the G3 have been chosen 
according to the cosmological constraints in Refs.~\cite{Peirone:2019aua} and~\cite{Peirone:2017vcq}, respectively.}
\label{fig:EERVepsilon} 
\end{figure}


In Fig.~\ref{fig:muLvsmuNL}, we show the time evolution of the non-linear effective gravitational coupling compared to 
the linear one for both the GGC and the G3 models. We note that the matter overdensity for the GGC case enters the 
Vainshtein radius approaching the GR solution before the G3 model does. The crossing time of the Vainshtein radius can 
be extrapolated from Fig.~\ref{fig:EERVepsilon} where we show the evolution of $(R/R_{\rm V})^3$. It occurs when 
$R/R_{\rm V}=1$ and for the GGC it is at $a=0.24$ and for the G3 at $a=0.47$.

The final stage of the collapse is virialization. The collapse stops when the system reaches the equilibrium and thus 
satisfies the Virial theorem. The latter states that, for a stable, self-gravitating, spherical distribution, the total 
kinetic energy of the object ($T$) and the total gravitational potential energy ($U$), satisfy the relation:
\be\label{virialtheorem}
 T + \f{1}{2}U = 0\,,
\ee
where the kinetic energy during the collapse for a top-hat profile is 
\be
 T \equiv \f{1}{2}\int\mathrm{d}^3x \,\rho_{\rm m} \mbox{v}^2 = \f{3}{10}M\dot{R}^2\,,
\ee
 and the total potential energy is \cite{Kimura:2010di} 
\ba
 U & \equiv & -\int \mathrm{d}^3x \,\rho_{\rm m}({\bf x}) \,{\bf x} \cdot \nabla \Psi \nn \\
 & = & \f{3}{5}\l(\dot{H}+H^2\r)MR^2 - 
       \f{3}{5}G_{\rm N}\mu^{\rm NL}\f{M}{R}\delta M\,.
\ea

Let us note that the energy conservation is not strictly satisfied for a time-dependent dark energy or modified gravity 
model \cite{Schmidt:2009yj}. Thus, we choose the virialization time $a_{\rm vir}$ such that the conservation 
relation~(\ref{virialtheorem}) is satisfied. We can then define the virial overdensity as
\be
 \Delta_{\rm vir} \equiv \f{\rho_{\rm vir}}{\rho_{\rm collapse}} = 
 \l[1+\delta_{\rm m}(R_{\rm vir})\r]\l(\f{a_{\rm collapse}}{a_{\rm vir}}\r)^3\,.
\ee

In Tab.~\ref{tab:bestvalues} we list some relevant physical quantities such as $a_{\rm vir}$, $R_{\rm vir}/R_i$, 
$\Delta_{\rm vir}$ and $\delta_{\rm c}$ for a matter overdensity collapsing at the present time. $\delta_{\rm c}$ is 
the linear critical density contrast defined as the value of the linear $\delta_{\rm m}$ at the collapse when initial 
conditions are assumed such that the non-linear equation diverges at the collapse time. Knowing the initial overdensity 
$\delta_{\rm i}$ and its time derivative $\dot{\delta}_{\rm i}$ leading to collapse at a given time, one solves the 
linearised version of Eq.~(\ref{nonlinearoverdensityeq}) to obtain the linear critical overdensity $\delta_{\rm c}$. 
This is mathematically equivalent to solve Eq.~(\ref{eqn:linpert}) and to rescale $\delta_{\rm i}$ by the linear growth 
factor of the corresponding cosmological model.

\begin{figure}[t!]
\includegraphics[width=.46\textwidth]{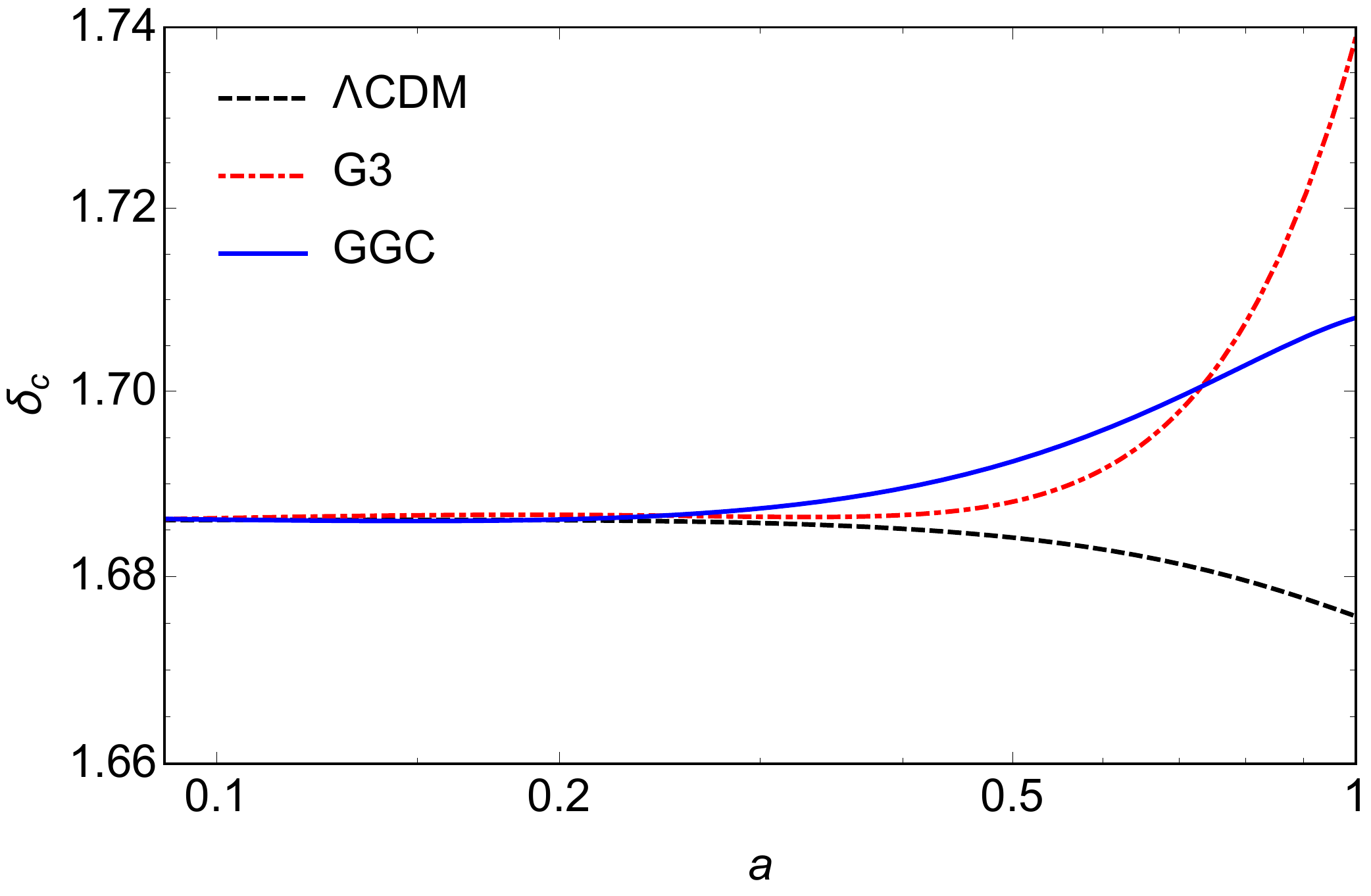}
\caption{Time evolution of $\delta_{\rm c}$ as function of the scale factor for the three cosmological models: 
$\Lambda$CDM (black dashed line), GGC (blue solid line) and G3 (red dot-dashed line).}
\label{fig:deltac}
\end{figure}

\begin{figure}[t!]
\includegraphics[width=.46\textwidth]{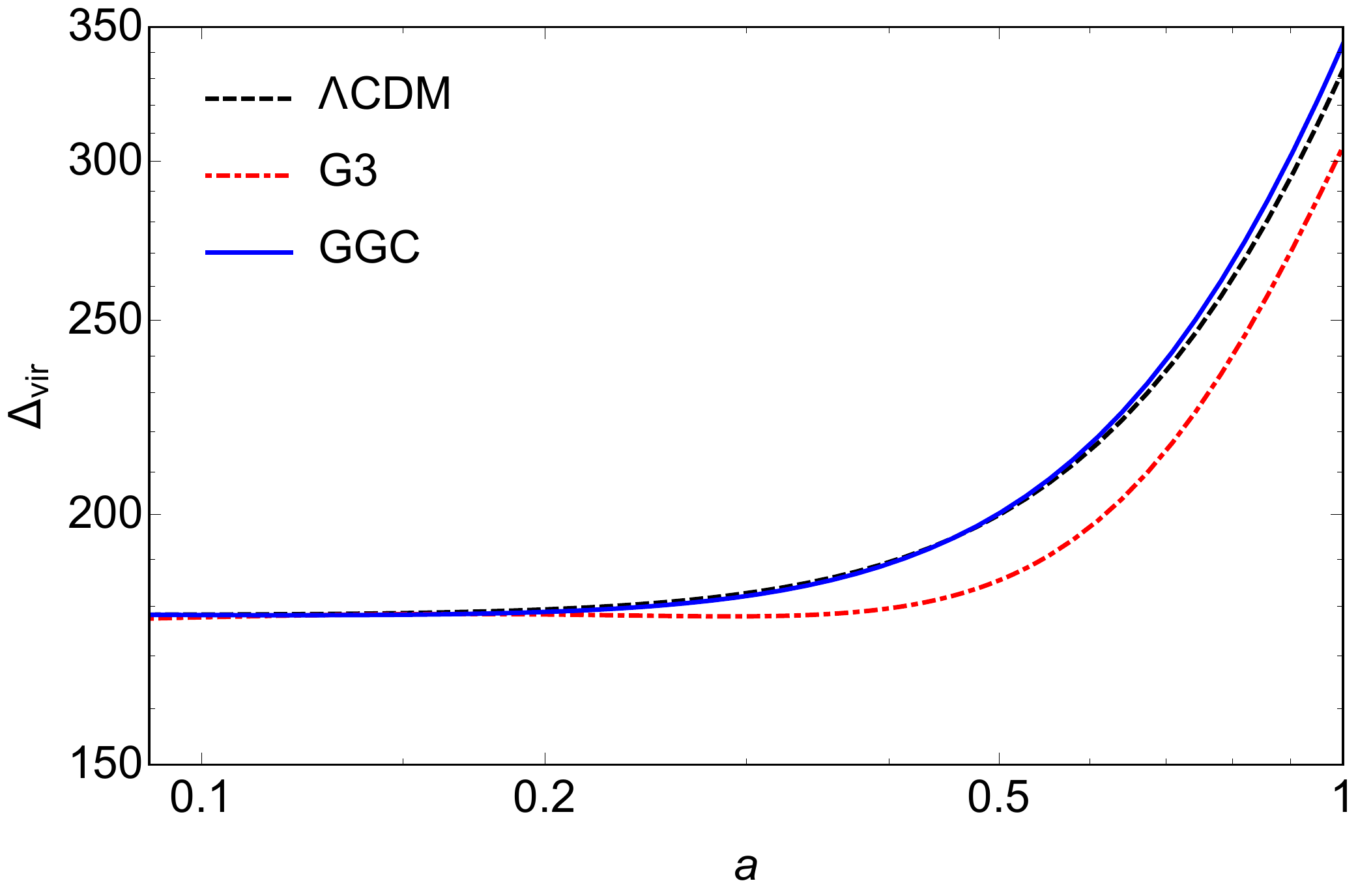}
\caption{Time evolution of $\Delta_{\rm vir}$ as a function of the scale factor for $\Lambda$CDM (black dashed line), 
GGC (blue solid line) and G3 (red-dot dashed line).}
\label{fig:deltavir}
\end{figure}

In Figs.~\ref{fig:deltac} and \ref{fig:deltavir} we show the evolution of $\delta_{\rm c}$ and $\Delta_{\rm vir}$ 
respectively as function of the scale factor. The critical density at early times approaches the value of the 
Einstein-de Sitter Universe, and for $a>0.2$ its time evolution differs in the three cosmological models, as the 
contribution of the cosmological constant and of the modifications of gravity become more important with time. In 
detail, while $\delta_{\rm c}(\Lambda{\rm CDM})$ decreases approaching the collapse at $a=1$, in the Galileon 
cosmologies the late time values of $\delta_{\rm c}$ are larger: $\delta_{\rm c}({\rm GGC})$ increases up to $1.708$ 
while $\delta_{\rm c}({\rm G3})$ is rather constant till $a\approx 0.4 $ and then it rapidly grows up to $1.738$. From 
Fig.~\ref{fig:deltavir} we see that the evolution of the virial overdensity for the $\Lambda$CDM and the GGC is 
approximately the same and $\Delta_{\rm vir}({\rm GGC})$ prefers slightly larger values for $a>0.5$. 
$\Delta_{\rm vir}({\rm G3})$ remains constant ($\approx 177.8$, the Einstein-de Sitter value) up to $a \approx 0.4$ and 
then increases remaining always smaller than $\Lambda$CDM and GGC.

We note that both the GGC and the G3 share in the Lagrangian the same form for the cubic term ($\propto X\Box \phi$), 
but in the latter the ghost condensate term $\propto X^2$ is not present. Although the modifications at both linear and 
non-linear regimes are driven by the cubic term, the inclusion of the $X^2$ term changes the evolution of the scalar 
field and that of the background quantities in such a way that all the relevant physical quantities we have 
investigated in this section show a significant modification with respect to the G3 model. Finally, we recall that 
these results are obtained using the best fit values for the parameters of each model from cosmological data on linear 
scales, thus according to data, the theoretical predictions we obtain are very close to what we can actually expect at 
non-linear level.

\section{Non-linear matter and lensing power spectra} \label{sec:nonlinmatterlensing}

Matter and lensing auto-correlation power spectra are two powerful tools to investigate the deviations from GR. One can 
resort to Einstein-Boltzmann codes to compute their predictions at linear scales~\cite{Bellini:2017avd} (see 
Ref.~\cite{Peirone:2019aua} for GGC). In order to extend such predictions on smaller scales one has to include 
non-linear corrections and screening mechanisms effects. These usually require a model by model implementation of the 
relevant equations in $N$-body codes 
\cite{Barreira:2013eea,Oyaizu2008,Schmidt2009,Zhao2011,Li2012,Brax2012a,Brax2012b,Baldi2012,Puchwein2013,Wyman2013,
Li2013,Llinares2014a,Llinares2014b,Llinares2018,Li2018,Llinares2020}.

Analytically, a formalism to calculate the non-linear matter power spectrum for wider classes of gravity models has 
been developed~\cite{Koyama:2009me} considering the closure approximation~\cite{Taruya:2007xy} with applications to DGP 
and $f(R)$ gravity models; or another approach~\cite{Cataneo:2018cic} is the one which extends the reaction method 
\cite{Mead:2016ybv} using the halo model. Alternatively, a parameterization based on  spherical collapse computations 
capturing  the non-linear MG effects on structure formation  has been recently proposed and implemented in an $N$-body 
code \cite{Hassani:2020rxd}.

In this work the goal is to have a glimpse into the phenomenology associated with the screening effects on the matter 
and lensing power spectra, leaving for a future work a more detailed investigation. In this regard, we will use the 
predictions from linear cosmological perturbation theory and incorporate the screening effects in a phenomenological 
fashion~\cite{Alonso:2016suf,Fasiello:2017bot,Reischke:2018ooh}.  We will model the small-scale limit to GR through a 
direct dependence on the screening scale in the matter power spectrum, as follows:

\begin{eqnarray}\label{nonlinearpk}
P_{GGC}^{\rm NL}(k,z) & = & \f{P_{\Lambda CDM}^{\rm NL}(k,z)}{P_{\Lambda CDM}^{\rm L}(k,z)} 
\left\{P_{\Lambda CDM}^{\rm L}(k,z)  \nn\r.\\
& + & \l.[P_{GGC}^{\rm L}(k,z)-P_{\Lambda CDM}^{\rm L}(k,z)]e^{-\l(\f{k}{k_s}\r)^2}\right\}\,,
\end{eqnarray}
where $k_{\rm s}$ is the screening scale. 
The linear power spectrum of $\Lambda$CDM and GGC are respectively obtained from the Einstein-Boltzmann solver CAMB 
\cite{Lewis:1999bs} and EFTCAMB \cite{Raveri:2014cka,Hu:2014sea}. The non-linear matter power spectrum for $\Lambda$CDM 
is obtained using the prescription in Ref.~\cite{Smith2003}. The cosmological parameters are the same for $\Lambda$CDM 
and GGC in Eq.~(\ref{nonlinearpk}), in particular they are those of GGC. The Eq.~(\ref{nonlinearpk}) recovers 
$P_{GGC}^{\rm L}(k,z)$ in the limit $k\ll k_{\rm s}$ and $P_{\Lambda CDM}^{\rm NL}(k,z)$ in the regime $k\gg k_{\rm s}$.

The value of the screening scale is strictly related to the specific model under consideration. From $N$-body 
cosmological simulations in the G3 model one finds that $k_{\rm s}=0.1 \,h\,{\rm Mpc}^{-1}$ at the present time 
\cite{Barreira:2013eea}. For the GGC model, $N$-body simulations do not exist, therefore we will present our results 
for four values of $k_{\rm s}$ in order to quantify the relevance of this parameter. 
They are $k_{\rm s}=0.05,0.1,0.5,1\, h\,{\rm Mpc}^{-1}$ and will serve to show the phenomenology of GGC at these scales 
and provide theoretical predictions to be then compared to accurate $N$-body simulations once they are available. 
We guess that the more reliable results for the GGC will be those with a screening scale larger than that of the G3 at 
the present time. That is because we find that the Vainshtein radius at the present time for a point source for GGC is 
slightly smaller than that in the G3 model, therefore we expect $k_{\rm s}({\rm GGC})>k_{\rm s}({\rm G3})$ at $z=0$. 
Note, though, that $k_{\rm s}({\rm GGC})\gtrsim k_{\rm s}({\rm G3})$ only very recently, while for the majority of the 
cosmic history, $k_{\rm s}({\rm GGC})\lesssim k_{\rm s}({\rm G3})$, as it can be easily seen from the evolution of the 
Vainshtein radius in Figs.~\ref{fig:Vradius} and~\ref{fig:EERVepsilon}. 
However, determining the time evolution of $k_{\rm s}$ is not an easy task and it is necessary to use $N$-body 
simulations for its accurate determination. Its time dependence is further confirmed in the case of the G3 model in 
Ref. \cite{Barreira:2013eea} using $N$-body simulations. For the G3 model the screening scale is 
$k_{\rm s}\simeq 0.1\,h\,{\rm Mpc}^{-1}$ at $a=1$ and $k_{\rm s}\simeq 0.3-0.4\,h\,{\rm Mpc}^{-1}$ at $a=0.6$. 
This is a clear indication that a constant $k_{\rm s}$ might just be a first approximation. 
Therefore to avoid introducing further phenomenological approaches, we prefer to consider the screening scale 
$k_{\rm s}$ constant in time, in agreement with current literature on the subject 
\cite{Alonso:2016suf,Fasiello:2017bot,Reischke:2018ooh}. While this approach does have a marginal effect on the study 
of the matter power spectrum, it might have relevance in the computation of the lensing power spectrum, as it requires 
the knowledge of the time evolution of both the matter power spectrum and the screening scale.

\begin{figure}[!t]
 \centering
 \includegraphics[width=0.53\textwidth]{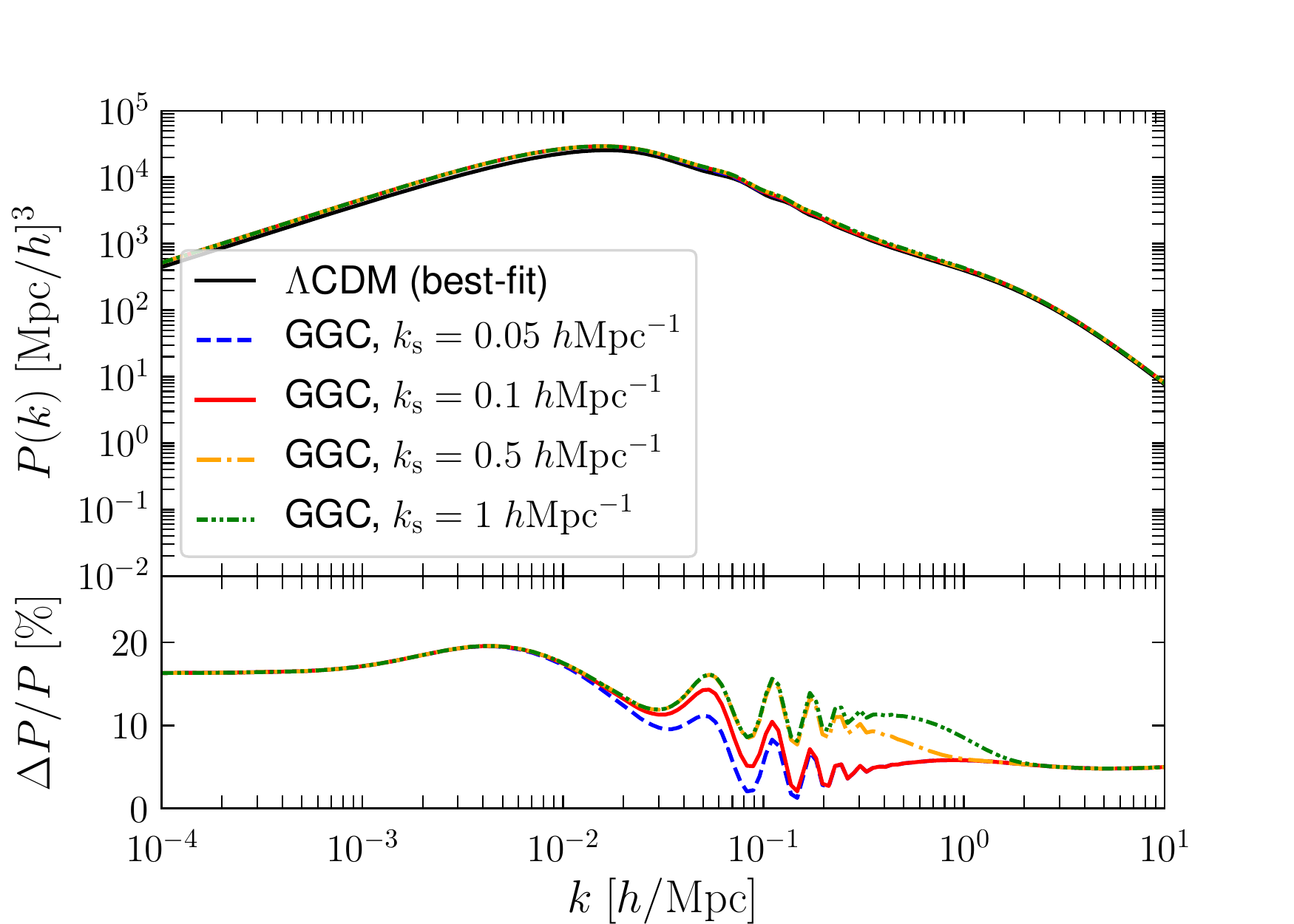}
 \includegraphics[width=0.53\textwidth]{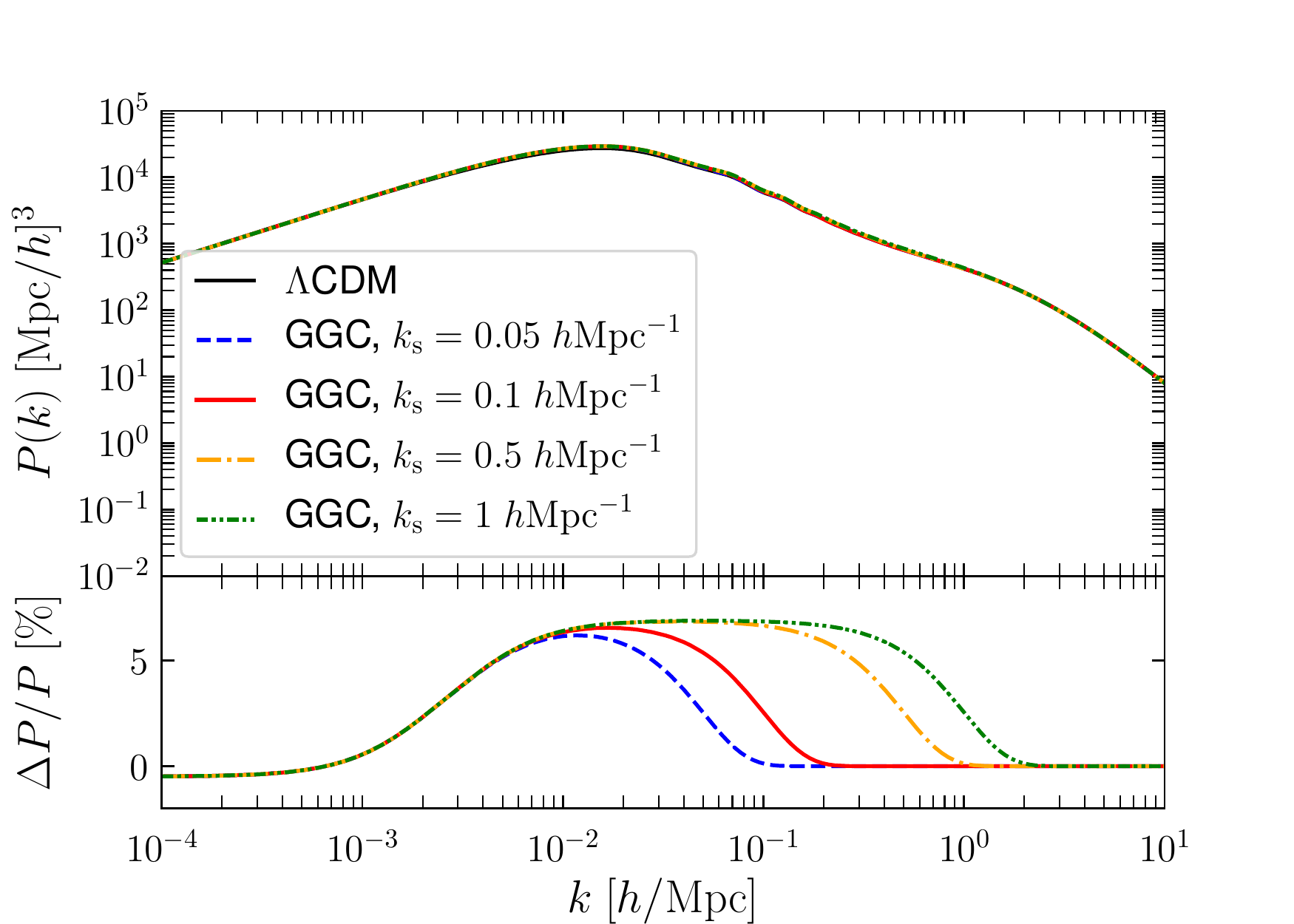}
 \caption{Non-linear matter power spectra as a function of $k$ at $z=0$ of the $\Lambda$CDM model (solid black line) 
 and of the GGC model, and non-linear matter power spectra percentage relative difference of the GGC model with respect 
 to $\Lambda$CDM. For the GGC we show the results for different screening scales $k_{\rm s}$: $0.05\,h\,{\rm Mpc}^{-1}$ 
 (dashed blue line), $0.1\,h\,{\rm Mpc}^{-1}$ (solid red line), $0.5\,h\,{\rm Mpc}^{-1}$ (dot-dashed orange line) and 
 $1\,h\,{\rm Mpc}^{-1}$ (dashed-dot-dotted green line). 
 The top figure compares the $\Lambda$CDM and the GGC models with the respective best-fit parameters 
 (Tab.~\ref{tab:bestvalues}), while the bottom one compares the GCC model with $\Lambda$CDM having the same 
 cosmological parameters of GGC.}
 \label{fig:Pk}
\end{figure}

In Fig.~\ref{fig:Pk} we present the non-linear matter power spectrum at $z=0$ for $\Lambda$CDM and the GGC model for 
the four screening scales discussed above and the relative difference $\Delta P/P$ in the bottom panel, where 
$\Delta P = P^{\rm GGC}(k)-P^{\Lambda{\rm CDM}}(k)$ and $P=P^{\Lambda{\rm CDM}}$. The top figure compares the two 
models having their best-fit parameters, while the bottom figure compares the GGC with the $\Lambda$CDM having the same 
cosmological parameters of the GGC. When comparing the two models with their best-fit parameters, on large scales, 
$k\lesssim 10^{-3}\,h\,{\rm Mpc}^{-1}$, the difference is smaller than 20\% between the two models due to the different 
normalizations of the spectra and a different behaviour of linear perturbations. The difference slightly increases on 
intermediate scales, up to 20\% and then decreases to approximately 5\% on small scales where the Vainshtein screening 
takes place. The exact scale depends on the value of the screening scale, $k_{\rm s}$. A small value of the latter 
induces a suppression of power on larger scales (small $k$) with respect to a larger value of $k_{\rm s}$. This is 
indeed evident when comparing the results for 
$k_{\rm s}=0.1\,h\,{\rm Mpc}^{-1}$ and $k_{\rm s}=0.5\,h\,{\rm Mpc}^{-1}$: 
a factor of 5 in the screening scale translates into about a factor of two in the scale where one would approximately 
recover the $\Lambda$CDM limit. The main difference in changing the screening scale is given by the scale at which the 
screening starts to be important. After reaching the maximum, for small values of the screening scale, the model looses 
power relatively fast, while large values of the screening scale lead to a slower decline of the power. All the GGC 
models, by construction, lead to the same plateau, which differs from zero as the GGC and $\Lambda$CDM models do not 
share the same cosmological parameters and as such it is not expected that the $\Lambda$CDM limit of the GGC power 
spectrum at non-linear scale coincides with that of $\Lambda$CDM with best-fit parameters. Because of this we also note 
that oscillations in the relative difference appear which originate from the baryon acoustic oscillations (BAO) 
signature imprinted on the matter power spectrum.

We recall that we made the comparison between GGC and $\Lambda$CDM respectively with their best-fit parameters with the 
purpose of spotting differences which can be closer to what we can actually observe. On the contrary in the bottom 
panel of Fig.~\ref{fig:Pk}, we compare the GGC model with the $\Lambda$CDM one having the same cosmological parameters 
of the GGC. In this case any difference we spot can be traced back to modified gravity only. The GGC model is slightly 
suppressed with respect to $\Lambda$CDM on very large scales $k< 10^{-3}$ h/Mpc. This is due to modifications in the 
evolution of the linear perturbations. Then, on intermediate scales ($10^{-3}\,h\,{\rm Mpc}^{-1}<k<k_{\rm s}$) we see 
an increase of power of about 7\%-8\%. This is a consequence of a stronger gravity force in the GGC model which is 
given by Eq.~(\ref{eq:mulinear}). Approaching $k_s$ the GGC matter power spectrum  declines due to the screening 
effect. The larger is the screening scale, the longer is the plateau.  For $k\approx 2k_{\rm s}$ the model is fully 
screened reaching the $\Lambda$CDM limit, as expected. Note that when comparing models with the same cosmological 
parameters, the wiggles in the ratio disappear, as the position of the BAO wiggles coincide.

We now investigate the effects of the GGC signatures on the lensing power spectrum. The latter is defined as the 
integral along the line of sight of the matter power spectrum. As for the matter power spectrum, we need to take into 
account the effects of modifications to gravity and on smaller scales we have to include those of the screening 
mechanism. The lensing effect depends on the sum of the two gravitational potentials, $\Phi+\Psi$, and as discussed in 
Sections~\ref{sec:linperturb} and \ref{sec:nonlinperturb} any departure form GR in the lensing equation can be included 
in the phenomenological function $\Sigma$. For GGC $\Phi=\Psi$ even on non-linear scale, so that $\Sigma=\mu$. In the 
following analysis we assume for $\Sigma$ the functional form \cite{Alonso:2016suf,Fasiello:2017bot,Reischke:2018ooh}
\be\label{GGGscreening}
\Sigma(k,z) = 1+\l(\Sigma^{\rm L}(z)-1\r)\exp{\l[-\l(\f{k}{k_{\rm s}}\r)^2\r]}\,.
\ee

The expression used to evaluate the lensing power spectrum is \cite{Bartelmann2001}\footnote{ With respect to 
Ref.~\cite{Bartelmann2001}, in Eq.~(\ref{eqn:Pl}) we include the modification to gravity with the function $\Sigma$.}
\begin{equation}\label{eqn:Pl}
 P_{\kappa}(\ell) = \f{9H_0^4{\Omega_{\rm m}^{(0)}}^2}{4c^4}
                    \int_0^{\chi_{\rm H}}\f{W(\chi)^2\Sigma(\chi,k)^2}{a^2(\chi)}
                    P^{NL}\l[\f{\ell+1/2}{\chi},\chi\r]\mathrm{d}\chi\,,
\end{equation}
where $W(\chi)$ is a kernel describing the distribution in redshift of the sources, $P(k)$ is the matter power spectrum 
evaluated at the wave-number $k=(\ell+1/2)/\chi$ \citep{Loverde2008}, being $\chi$ the comoving distance. Finally, 
$\chi_{\rm H}$ represents the comoving distance of the horizon. The function $\Sigma$ depends on the scale $k$ and the 
time (here parameterized via $\chi$). Assuming there is no scale-dependent screening, the expression in 
Eq.~(\ref{eqn:Pl}) reduces to Eq.~(47) of Ref.~\cite{Pace2014} upon the following identification $\Sigma=1/F(a)$. Also 
note that, for simplicity, we assumed the sources to be fixed in redshift at $z_{\rm s}=2$. Distributing the sources in 
redshift will not change our conclusions qualitatively, but only slightly decrease the impact of the modifications.

\begin{figure}[!t]
 \centering
 \includegraphics[width=0.53\textwidth]{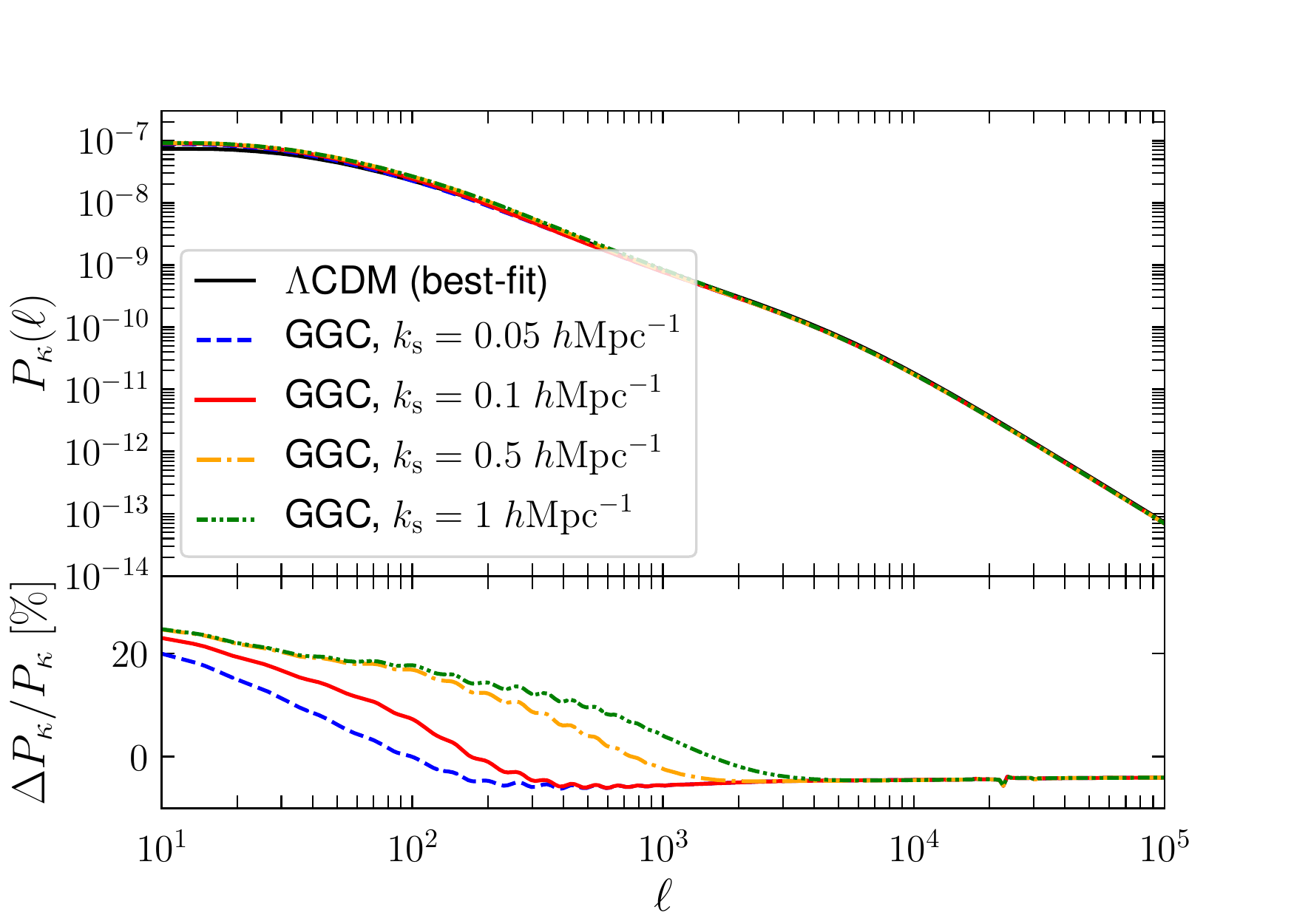}
 \includegraphics[width=0.53\textwidth]{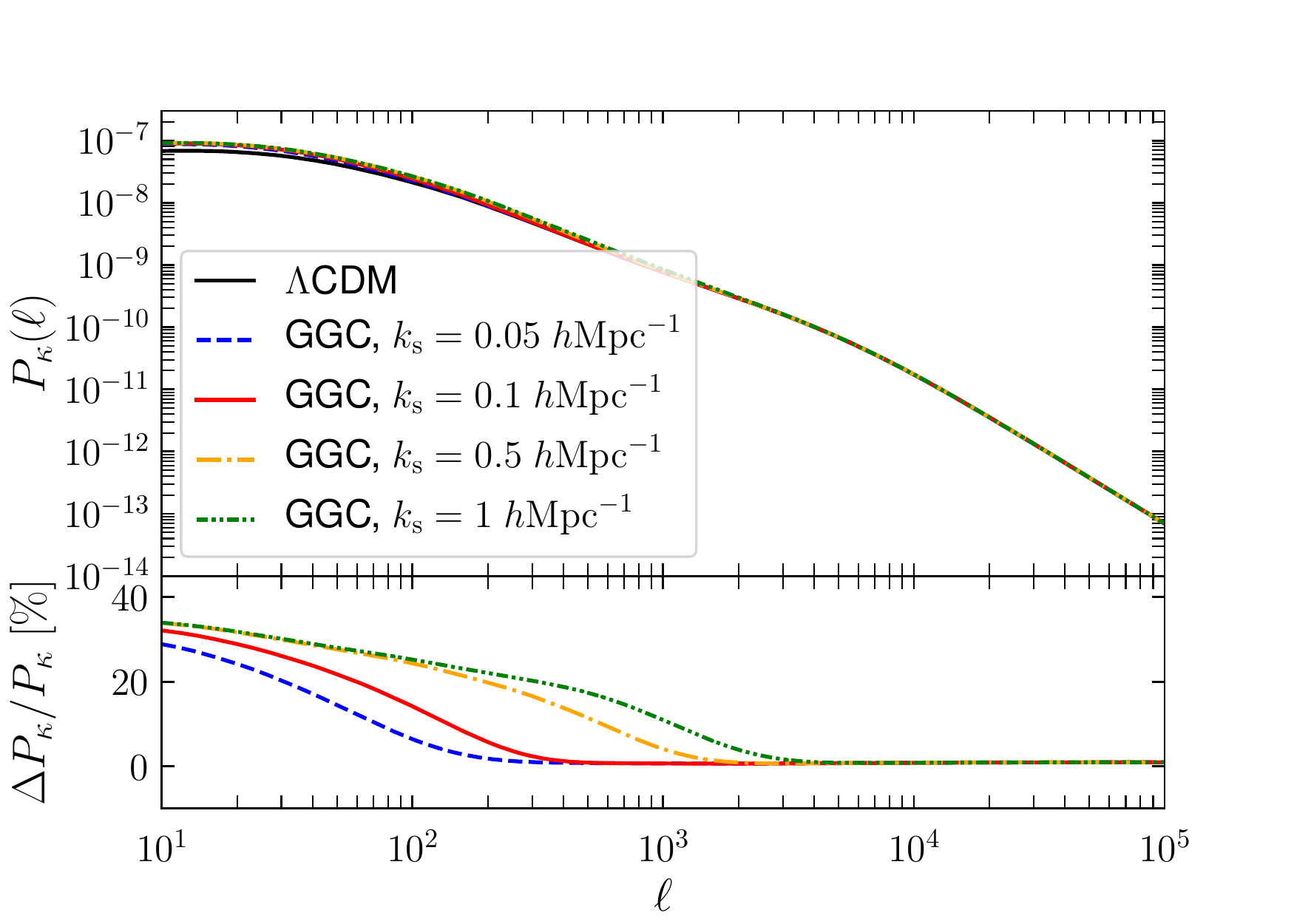}
 \caption{Non-linear lensing power spectra as a function of the multipole $\ell$ for sources at $z_{\rm s}=2$ for 
 $\Lambda$CDM (solid black line) and for the GGC model; and non-linear lensing power spectra percentage relative 
 difference of the GGC model with respect to $\Lambda$CDM. For the GGC, we show the results for different screening 
 scales, $k_{\rm s}=0.05, 0.1, 0.5, 1\,h\,{\rm Mpc}^{-1}$ using the same colour- and line-style of Fig.~\ref{fig:Pk}. 
 The top figure considers the $\Lambda$CDM and the GGC models with the respective best-fit parameters 
 (Tab.~\ref{tab:bestvalues}), while the bottom one assumes that the $\Lambda$CDM model has the same cosmological 
 parameters of the GGC model.}
 \label{fig:lensing}
\end{figure}

We show in Fig.~\ref{fig:lensing} the results for the non-linear lensing power spectrum in both the $\Lambda$CDM and 
the GGC scenarios. The latter is given for different screening scales. As for the matter power spectrum, we compare the 
two models considering both their best-fit parameters (upper figure) or when they have the same cosmological parameters 
(lower figure). 
Because the screening affects small scales (high-$\ell$), both models look almost the same in this regime and the 
multipole where this happen depends, obviously, on  the screening scale $k_{\rm s}$. Larger differences are restricted 
to small-$\ell$. 

Let us start discussing the case in which the two models are characterized by their best-fit parameters (upper panel in 
Fig.~\ref{fig:lensing}). The GGC lensing power spectrum is enhanced at $\ell=10$ with respect to $\Lambda$CDM up to 
25\% depending on the particular screening scale. This originates from the parameter $\Sigma(k,z)$ in 
Eq.~(\ref{eqn:Pl}). Due to the screening, the power decreases linearly until it reaches a plateau for large $\ell$. The 
rate at which the GGC approaches the plateau is faster for smaller $k_{\rm s}$, as the screening takes place at larger 
scales. For the smaller values of the screening scale, it is reached at $\ell\approx 200$, while for 
$k_{\rm s}=1\,h\,{\rm Mpc}^{-1}$ it is at $\ell\gtrsim 3000$. For the lensing power spectrum, a change in $k_{\rm s}$ 
of a factor of ten changes the scale at which the spectrum of the GGC approaches the plateau roughly by the same amount 
(see, for example, the behaviour for $k_{\rm s}=0.1\,h\,{\rm Mpc}^{-1}$ and $k_{\rm s}=1\,h\,{\rm Mpc}^{-1}$). 
The 5\% suppression in the plateau is due to the different cosmological parameters mostly related to the 
$\Omega_{\rm m}^{(0)}$ pre-factor in Eq.~(\ref{eqn:Pl}) which is higher in the $\Lambda$CDM best-fit case.

Let us now consider the case in which both $\Lambda$CDM and GGC share the same cosmological parameters (bottom panel in 
Fig.~\ref{fig:lensing}). The effects of modified gravity are more pronounced at small $\ell$ and this is due to an 
higher difference in $\Sigma$ with respect to the previous case and it can be up to 35\%, a $\sim 10\%$ larger than the 
best-fit case. At intermediate scale the ripples disappear as expected due to the fact that the matter power spectrum 
does not show them any more. Finally, when the power spectrum reaches the plateau the discrepancy observed in the 
previous case disappears as both the power spectrum and $\Sigma$ are in the $\Lambda$CDM limit and the pre-factor in 
Eq.~(\ref{eqn:Pl}) (i.e. $\Omega_{\rm m}^{(0)}$) is the same.

\section{Mass function}\label{sec:massfunction}
In this section we investigate the effects of the GGC model on the abundance of halos. To this purpose we use the Sheth 
\& Tormen mass function~\cite{Sheth1998,Sheth2001,Sheth2002,Murray2013} 
\ba
\f{\mathrm{d}n}{\mathrm{d}M} & = & -\sqrt{\f{2\tilde{a}}{\pi}}A
\l[1+\l(\f{\tilde{a}\,\delta_{\rm c}^2}{D^2\sigma_M^2}\r)^{-p}\r]
\f{\rho_{\rm m}}{M^2}\f{\delta_{\rm c}}{D\sigma_M}\nn\\
&& \times\,\f{\mathrm{d}\ln{\sigma_M}}{\mathrm{d}\ln{ M}}
\exp{\l(-\f{\tilde{a}\,\delta_{\rm c}^2}{2D^2\sigma_M^2}\r)}\,,
\ea
where $\tilde{a}=0.707$~\footnote{We changed the commonly adopted notation to avoid confusion with the scale factor 
$a$.}, $p=0.3$, $A=0.2162$, $\delta_{\rm c}$ is the linear critical density contrast derived in 
Section~\ref{Sec:sphericalcollapse}, $D=\delta^{\rm L}_{\rm m}/\delta^{\rm L}_{\rm m}(a=1)$ is the linear growth 
factor, and $\sigma_M$ is the variance of the linear matter power spectrum defined as \cite{Press1974}
\be
\sigma_M^2 = \f{1}{2\pi^2}\int^\infty_0 \mathrm{d}k\, k^2W^2(kR)\,P^L(k)\,,
\ee
where the window function is defined as
\be
W(kR) = 3\f{\sin(kR)-kR\cos(kR)}{(kR)^{3}}\,,
\ee
being $R$ the comoving radius enclosing the mass $M=\tfrac{4\pi}{3}\rho_{\rm m}R^3$. The window function represents the 
Fourier transform of the top-hat function in the space configuration. We also compute the number density of objects 
above a given mass at a chosen $z$ as:
\be
n(>M) = \int^\infty_M\f{\mathrm{d}n}{\mathrm{d}M^{\prime}}\mathrm{d}M^{\prime}\,.
\ee
\begin{figure}[t!]
 \centering
 \includegraphics[width=0.5\textwidth]{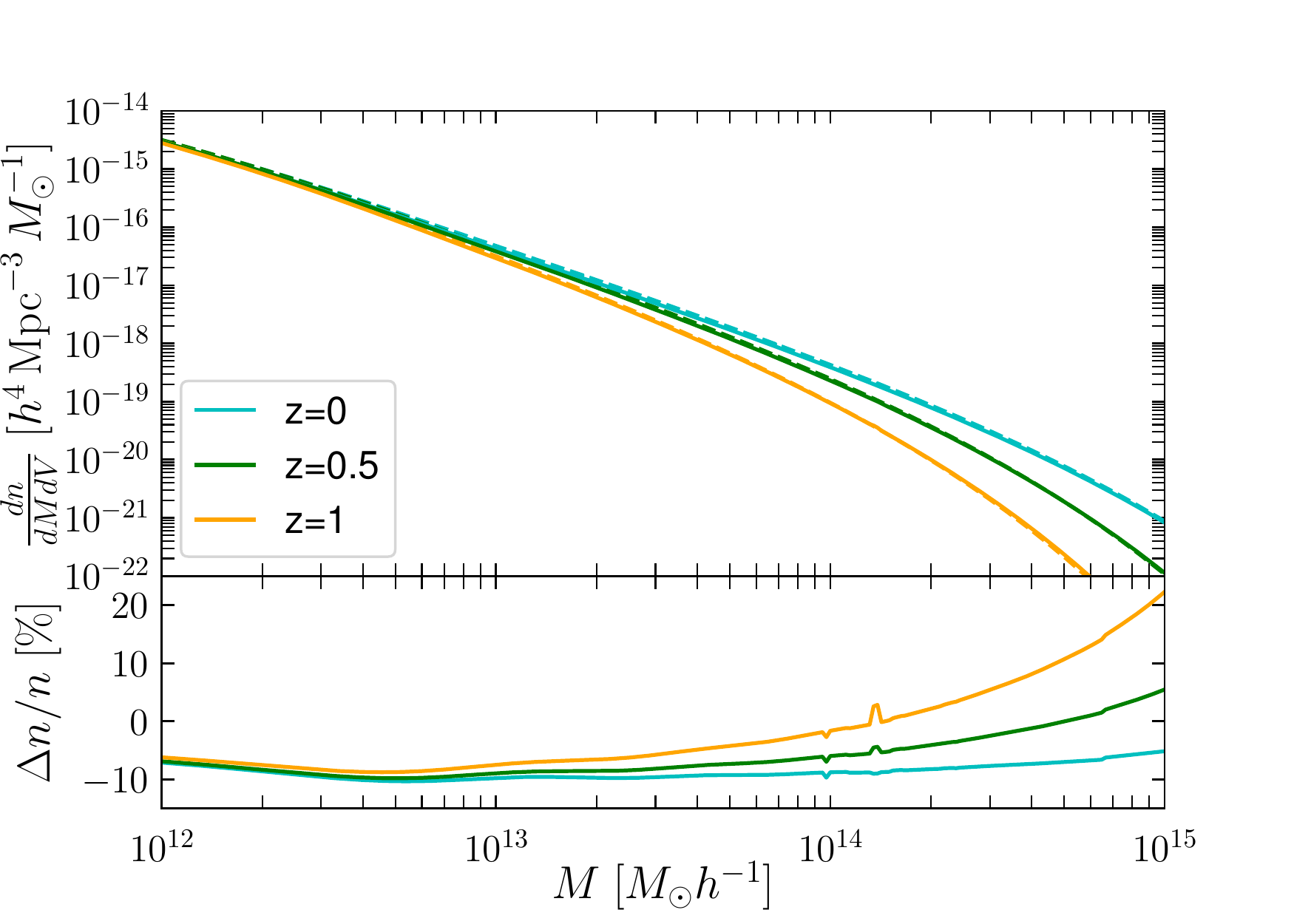}
 \includegraphics[width=0.5\textwidth]{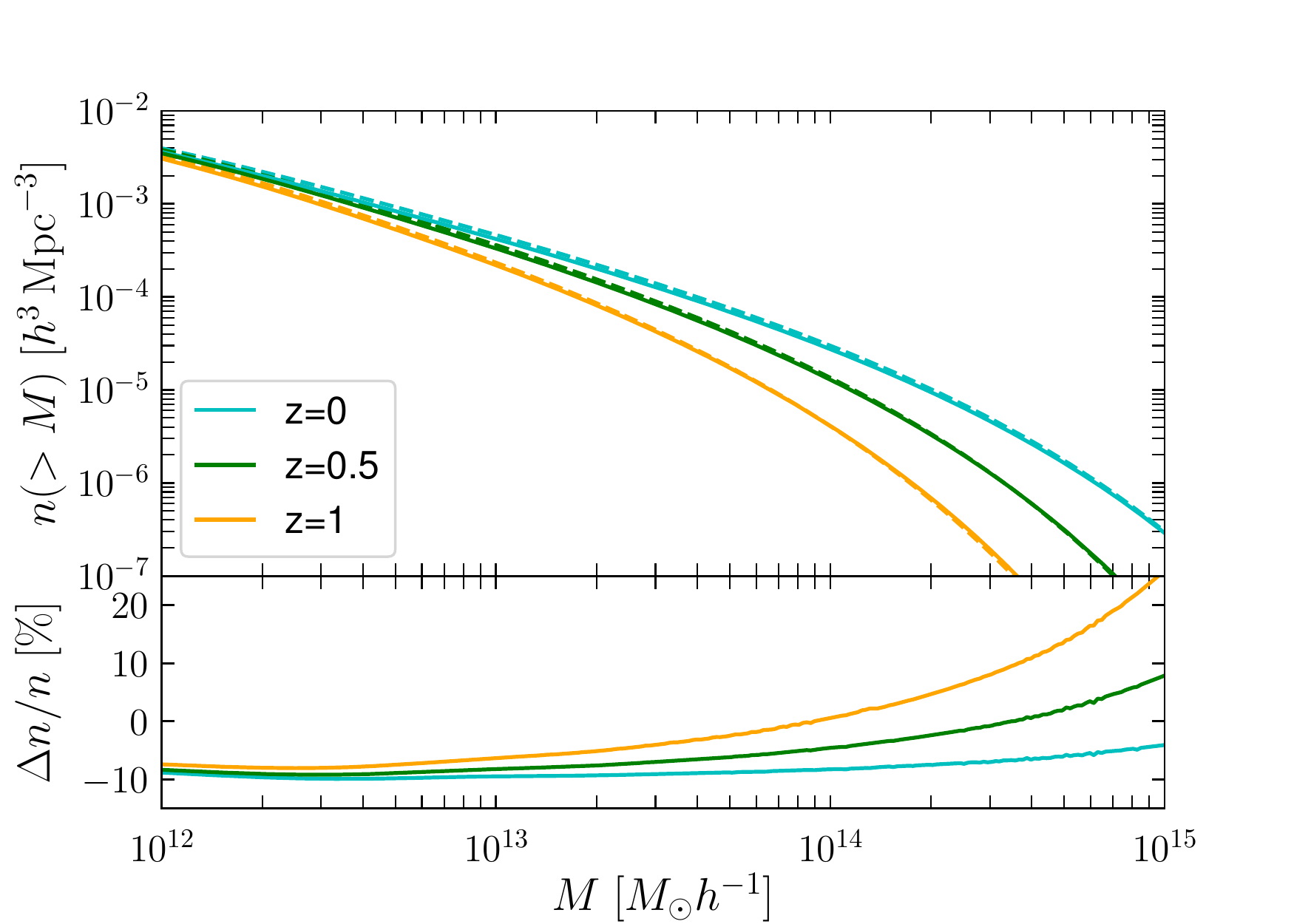}
 \caption{Top panel: differential mass function as a function of the halo mass $M$ at $z=0, 0.5, 1$, as shown in the 
 labels. Solid lines refer to the GGC model, while dashed lines to $\Lambda$CDM. Bottom panel: cumulative mass function 
 as a function of the halo mass for the same set of redshifts. The cosmological and models parameters of $\Lambda$CDM 
 and GGC models are the best-fit ones in Tab. \ref{tab:bestvalues}.}
 \label{fig:mf_bestfit}
\end{figure}

\begin{figure}[t!]
 \centering
 \includegraphics[width=0.5\textwidth]{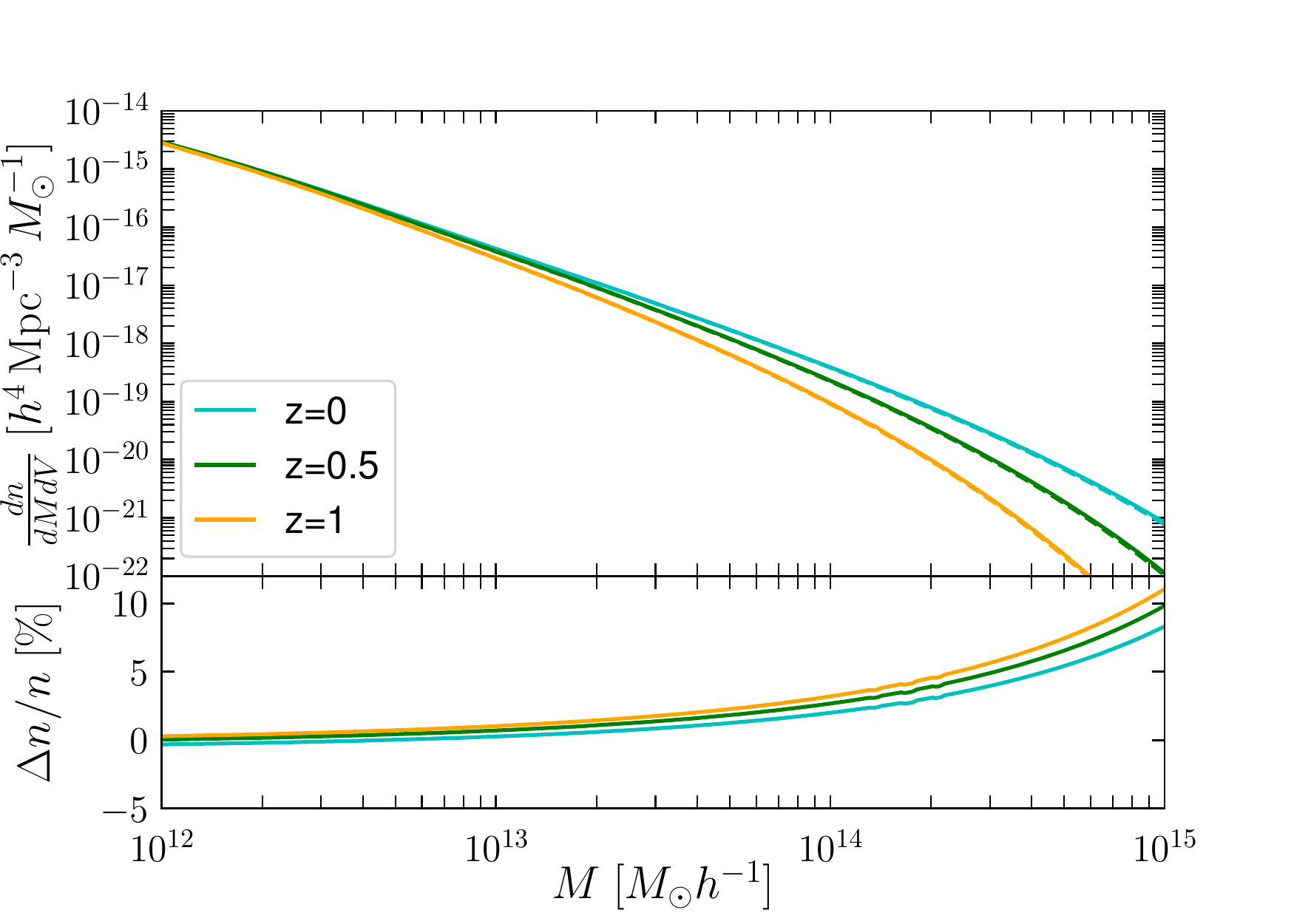}
 \includegraphics[width=0.5\textwidth]{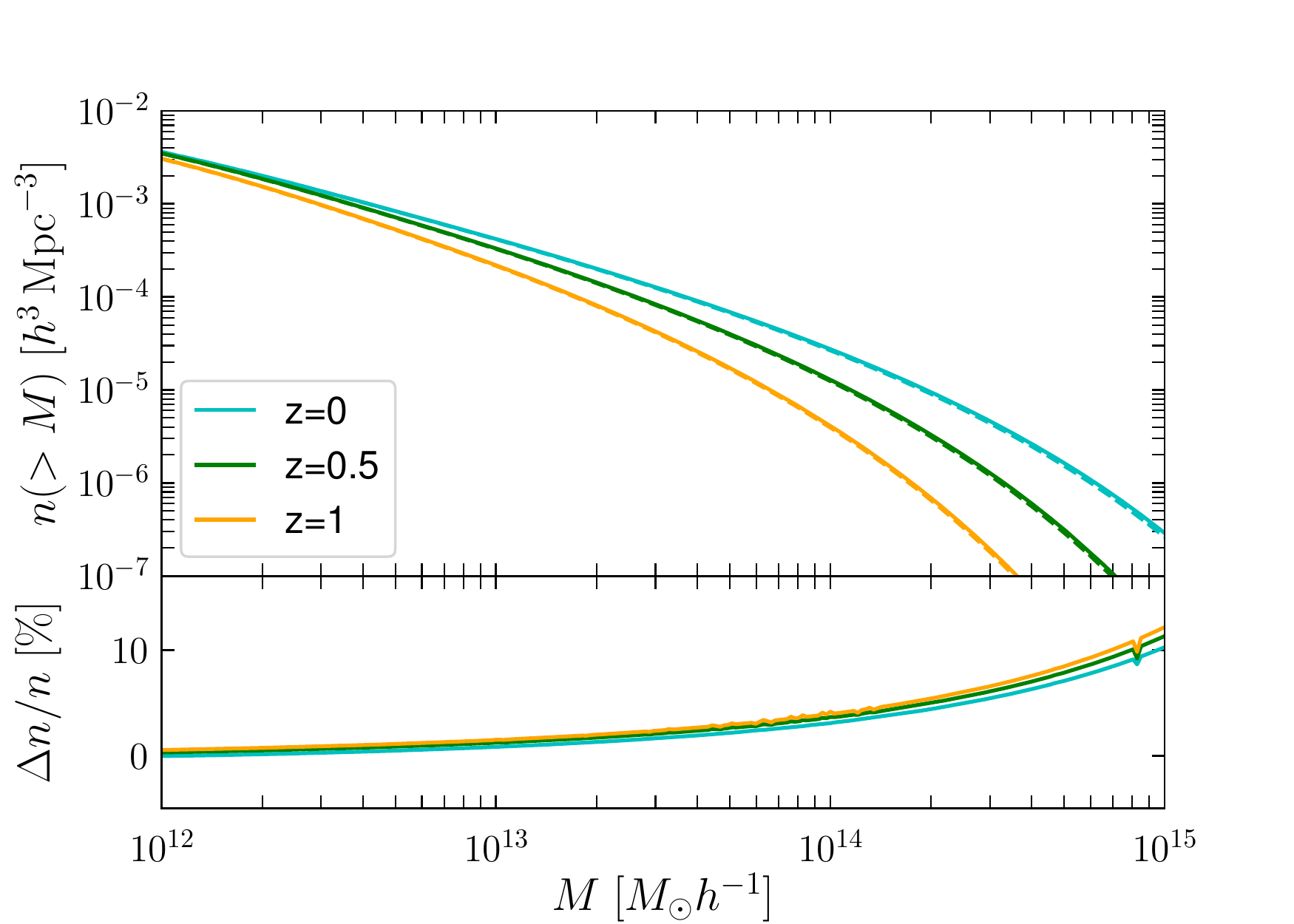}
 \caption{Same as in Fig.~\ref{fig:mf_bestfit}, but now the $\Lambda$CDM shares the same cosmological parameters of the 
 GGC model.}
 \label{fig:mf_LCDM_GGC}
\end{figure}

In the expression for the mass function, whilst we keep the same constants ($A$, $\tilde{a}$ and $p$) for both the GGC 
and $\Lambda$CDM, the physical parameters ($\delta_{\rm c}$, $\sigma_{M}$, $D$) are consistently computed for each 
model.

We present the results in Fig.~\ref{fig:mf_bestfit} and in Fig.~\ref{fig:mf_LCDM_GGC}, where we show the differential 
and the cumulative mass functions. We also consider the relative difference of the GGC model with respect to 
$\Lambda$CDM, i.e. $\Delta n/n = (n^{\rm GGC}-n^{\Lambda{\rm CDM}})/n^{\Lambda{\rm CDM}}$. We select three redshifts 
$z=0, 0.5, 1$ and we consider halo masses ranging from $10^{12}\,M_{\odot}\,h^{-1}$ (galactic scales), until 
$10^{15}\,M_{\odot}\,h^{-1}$ (cluster scales) to better assess the effects of GGC at different mass scales and 
redshifts. In Figs.~\ref{fig:mf_bestfit} and \ref{fig:mf_LCDM_GGC} we compare GGC and $\Lambda$CDM having their 
best-fit parameters and the same cosmological parameters, respectively.

When comparing the models using their best-fit parameters, for low masses we observe a decrease of about 10\% in the 
GGC model with respect to $\Lambda$CDM, regardless of the chosen redshift. At $z=0$, the lack of objects is rather 
constant over two orders of magnitudes in mass in the two models. The number of objects become more similar towards 
high masses, but the GGC model still shows a few percent less objects than $\Lambda$CDM. At higher redshifts, the 
differences observed at low masses get smaller going towards $M \approx 10^{14}\,M_{\odot}\,h^{-1}$ for $z=0.5$ and 
$M \approx 3 \times 10^{13}\,M_{\odot}\,h^{-1}$ for $z=1$, respectively, but for higher masses they become more 
prominent and it is where the two models differ the most in the predictions of the number of halos.

What noticed is the typical behaviour of models beyond $\Lambda$CDM. The reason is that the exponential suppression in 
the halo mass function (and, as a consequence, also in the cumulative mass function) is more important at high masses 
and redshifts \cite{Sheth1998,Sheth2001,Sheth2002}. 
The GGC model predicts an excess of objects in the high-mass tail: for $z=0.5$ and $z=1$ we find, respectively, up to 
10\% and 20\% more objects. The differences between GGC and $\Lambda$CDM models using the best-fit parameters are 
caused by the substantially different evolution of the linear critical overdensity $\delta_{\rm c}$ (see 
Fig.~\ref{fig:deltac}) and of the mass variance $\sigma_M$ as the linear matter power spectra in the two cosmological 
models differ by almost 20\% on large scales ($k\approx 10^{-4}\,h\,{\rm Mpc}^{-1}$) and 5\%-7\% on small scales 
($k\approx 2\,h\,{\rm Mpc}^{-1}$). More in detail, the matter power spectrum in the GGC model is higher than the  
$\Lambda$CDM one, and the same happens for the variance $\sigma_M$. This implies that while a higher value of 
$\delta_{\rm c}$ leads to a suppression in the mass function, a higher value of $\sigma_M$ instead leads to an 
enhancement. 
At low redshifts, as the differences due to $\delta_{\rm c}$ are bigger than those of $\sigma_M$ and 
$\delta_{\rm c}({\rm GGC})>\delta_{\rm c}(\Lambda{\rm CDM})$, we have a suppression in the number of objects; at high 
redshifts $\delta_{\rm c}({\rm GGC})\gtrsim\delta_{\rm c}(\Lambda{\rm CDM})$ and the major effect is due to the mass 
variance $\sigma_M$, hence we observe an increase of the relative mass function at high masses. 
At low masses, the exponential term contributes less to the overall picture with respect to the other terms and the 
decrease in the number of objects is due to the latter.

Instead, when we compare the two models using the same cosmological parameters, they have the same behaviour on small 
masses and the GGC, due to a higher clustering, predicts more massive halos than the $\Lambda$CDM. With respect to the 
best-fit case, differences are slightly smaller (up to $10\%$) and we do not find a significant dependence with 
redshift. This is in agreement with our previous findings about the matter and lensing power spectra, as small masses 
pick up the linear part of the matter power spectrum.

From the observational side it would be of interest to compare the predicted halo mass function for the GGC model with 
the data obtained in Ref.~\cite{Abdullah:2020qmm} from a sub-sample of 843 clusters (SelFMC) in the redshift range 
$0.01 \leq z \leq 0.125$ with virial masses of $M \geq 0.8 \times 10^{14}\,h^{-1}\,M_{\odot}$ from the GalWCat19 
catalogue \cite{Abdullah:2019hsx}. 
As the catalogue is complete in the mass range of $10^{14}<M/M_\odot<10^{15}$, where the GGC predictions largely 
differ from $\Lambda$CDM, it is possible to better assess the influence of modifications of gravity. An extension of 
the analysis to smaller masses or different redshift ranges, would require to weigh the observed mass function with a 
selection function $S(D)$, where $D$ is the comoving distance of the cluster. For more details about the procedure, we 
refer the reader to \cite{Abdullah:2020qmm}. 
Let us open a parenthesis about the impact of any assumption on the underlying cosmological model made to derive the 
data. For example, since the catalogue includes only objects at small redshift, any influence due a different 
background evolution can be safely neglected  when considering the distance of the clusters and the cosmic volume 
spanned by the survey as they can be approximated with $c/H_0$ and $(c/H_0)^3$, respectively, where $c$ is the speed 
of light. However, in the process of mass calibration modifications of gravity might play a role. To this purpose, we 
can consider, for example, the mass-temperature relation used in \cite{Hjorth1998a,Hjorth1998b,Campanelli2012} which 
reads
\begin{equation}
 \tilde{M} = 1.5\times 10^{14}\,h^{-1}\,M_{\odot}\kappa_{\Delta}\frac{T_X}{{\rm keV}} \frac{1}{1+z}\,,
\end{equation}
where $\tilde{M}$ is the virial mass contained in a comoving radius $R_0^{\prime}=1.5\,h^{-1}\,{\rm Mpc}$ and $T_X$ 
the cluster X-ray temperature. The quantity $\kappa_{\Delta}$ depends on the virial overdensity $\Delta_{\rm vir}$ 
which can change in modified gravity cosmologies. In Fig.~\ref{fig:deltavir} we showed that 
$\Delta_{\rm vir}^{\rm GGC}\simeq \Delta_{\rm vir}^{\rm \Lambda CDM}$ for the best-fit parameter values over all the 
cosmic history relevant to this work. Hence, the dependence on the modified cosmological model is removed in the case 
under analysis. On the contrary this will not be the case for G3.

\section{Conclusions}\label{sec:conclusion}

In this work we studied the impact of non-linearities in the Galileon ghost condensate (GGC) model \citep{Kase:2018iwp} 
on the formation of spherical gravitationally bounded objects and we made theoretical predictions on the abundances of 
halos, non-linear matter and lensing power spectra. To spot key features we have compared the results with the 
standard cosmological scenario $\Lambda$CDM and another Galileon model, the cubic Galileon (G3) 
\citep{Deffayet:2009wt} which shares with the GGC the term in the Lagrangian $\propto X\Box\phi+X$ but differs for the 
$X^2$ term which is not present in the G3. The results presented in the analysis used the maximum likelihood values 
for the cosmological and model parameters obtained with Planck data in previous works. This is because we wanted to 
show the difference between the predictions of every model as close as possible to what we actually expect from 
observations.

We found that the predictions on the growth of structures and spherical collapse of the GGC model are quite different 
from those of the G3 and are closer to the $\Lambda$CDM ones but still with some peculiarities. To start with, the 
linear growth rate in G3 presents large enhancements with respect to the GGC, being the latter very close to 
$\Lambda$CDM. On non-linear scales, the presence of the Vainshtein screening mechanism which characterises both 
Galileon models changes the gravitational coupling felt by matter which, in both cases, is larger than that in 
$\Lambda$CDM. We noted that for a collapse taking place at the present time in the GGC model such gravitational 
coupling stays closer to $\Lambda$CDM than the G3 one. This is due to the fact that the former enters in the 
Vainshtein radius before. Indeed, during the collapse process, the Vainshtein radius of the GGC is always larger than 
that in the G3. Furthermore, we found that the turn-around phase for both Galileon models takes place slightly before 
than for $\Lambda$CDM and the virialization time follows the order G3, GGC and $\Lambda$CDM. Being G3 the first to 
reach virialization, the evolution of the virial overdensity for the Galileon models is completely different: while G3 
stays always below $\Lambda$CDM, the GGC closely follows the $\Lambda$CDM one and after $a \approx 0.5$ it is slightly 
enhanced. The evolution of the linear critical overdensity $\delta_{\rm c}$ shows again key features after $z \approx 
2$: in the $\Lambda$CDM scenario it decreases from the de-Sitter value to $\approx 1.675$ at present time; in the GGC, 
instead, it has the opposite behaviour, increasing its value up to $1.708$; finally the G3 grows as well but up to $a 
\approx 0.7$ it stays below the GGC and after it overcomes the GGC, reaching the present day value of $\approx 1.74$. 
These new features of the GGC can be addressed considering the inclusion of the $X^2$ term in the Lagrangian which 
makes the difference with respect to G3. This term changes remarkably the evolution of the Vainshtein radius and as 
such the physics associated to the formation of (non-linear) structures.

We employed a phenomenological approach to incorporate the screening mechanism in the computation of the non-linear 
matter and lensing power spectra. This is done by considering the gravitational couplings felt by matter and light, 
respectively, to have an explicit dependence on the screening scale, $k_{\rm s}$ such that when $k<k_{\rm s}$ it 
reduces to the linear GGC spectrum while when $k>k_{\rm s}$ they approach the $\Lambda$CDM behaviour. Because we do 
not know the screening scale of the GGC model in Fourier space, we computed the predictions for matter and lensing 
power spectra for four scales. This approach led us to show the phenomenology of GGC and provide theoretical 
predictions which, in the future, can be compared to accurate $N$-body simulations once they are available. We found 
that when comparing the models characterized by their best-fit parameters, the matter power spectrum on linear scales 
for GGC shows a difference with respect to $\Lambda$CDM smaller than $20\%$ and it decreases to a few percent on the 
smaller scales. The scale at which the matter power spectrum approaches a plateau on large $k$ depends on the 
screening scale. As expected, smaller screening scales suppress the matter power spectrum at larger scales. When we 
compare the models using the same base cosmological parameters the difference in the matter power spectrum persists 
only on intermediate scales $10^{-3} \, h\,{\rm Mpc}^{-1}<k<k_{\rm s}$ (up to $7\%-8\%$). In the lensing power 
spectrum, the relative difference between GGC and $\Lambda$CDM for the best-fit case, exceeds $20\%$ at small-$\ell$ 
and decreases at larger-$\ell$. The values of $k_{\rm s}$ we chose show that for the smaller value of 
$k_{\rm s}=0.05\,h\,{\rm Mpc}^{-1}$, the $\Lambda$CDM limit is reached at $\ell\approx 200$ and for the larger 
$k_{\rm s}=1\,h\,{\rm Mpc}^{-1}$ we found $\ell\approx 3000$. 
A similar behaviour characterizes the comparison when the same cosmological parameters are employed, but in this case 
the difference at small $\ell$ is 10\% larger. We then computed the mass function as a function of the halo mass 
following the Sheth \& Tormen model. We found that for the best-fit case at low masses the GGC model provides about 
10\% less objects with respect to $\Lambda$CDM, while at higher masses and higher redshift ($z>0.5$) it predicts about 
10\%-20\% more objects. This can be explained by the fact that the critical linear overdensity for GGC is larger 
than in $\Lambda$CDM and by a larger mass variance in the former. When the difference due to the different 
cosmological parameters is removed  we found that at small masses both models predict the same number of objects but 
at larger masses GGC predicts up to 10\% more objects than $\Lambda$CDM regardless of the redshift.

Finally, given the results presented in this paper, the GGC model shows very peculiar and measurable features which 
can definitely help in discriminating between GGC and $\Lambda$CDM. Whilst this work provides only a glimpse into the 
phenomenology of non-linear matter and lensing power spectra, less simplified methods can be employed, such as those 
in Refs. \cite{Koyama:2009me,Cataneo:2018cic,Hassani:2020rxd}, which we will consider in an upcoming work.

We further stress that a proper assessment and validation of our results can come with realistic $N$-body simulations, 
which are not affected by the necessary simplifications required for an analytical evaluation. Simulations will also 
allow to produce fitting formulae for the evolution of the non-linear matter power spectrum and improve the formalism 
of the spherical collapse model.

\section*{Acknowledgements}
\noindent
We thank Alberto Rozas-Fern{\'a}ndez, Bj{\"o}rn Malte Sch{\"a}fer and Shinji Tsujikawa for useful discussions.
The research of NF is supported by Funda\c{c}\~{a}o para a Ci\^{e}ncia e a Tecnologia (FCT) through the research 
grants UID/FIS/04434/2019, UIDB/04434 /2020 and UIDP/04434/2020 and by FCT project ``DarkRipple -- Spacetime ripples 
in the dark gravitational Universe" with ref.~number PTDC/FIS-OUT/29048/2017. FP acknowledges support from Science and 
Technology Facilities Council (STFC) grant ST/P000649/1 and the ERC Consolidator Grant \textit{CMBSPEC} (No.~725456) 
as part of the European Union's Horizon 2020 research and innovation program.

\bibliography{GGC.bbl}

\end{document}